\PassOptionsToPackage{table}{xcolor}

\documentclass[3p]{article}

\usepackage[hyperindex,breaklinks]{hyperref}

\def\urlprefix{}

\usepackage{todonotes}

\usepackage{url}

\usepackage{booktabs}
\usepackage{amsfonts}
\usepackage{subfigure}
\usepackage{amsmath}
\usepackage{acronym}
\usepackage{multirow}
\usepackage{array}
\usepackage{hanging}
\usepackage{xcolor}
\usepackage{xspace}
\usepackage{floatrow}
\usepackage[export]{adjustbox}
\usepackage{tikz}
\usepackage{epstopdf}
\usepackage{enumitem}

\usetikzlibrary{shapes,arrows,positioning}
\usetikzlibrary{arrows.meta}

\acrodef{HH}{household}
\acrodef{HC}{household cluster}
\acrodef{SLA}{statistical local area}
\acrodef{DZN}{destination zone}
\acrodef{CD}{census collection district}
\acrodef{WG}{working group}
\acrodef{ABS}{Australian Bureau of Statistics}
\acrodef{BITRE}{Australian Bureau of Infrastructure, Transport and Regional Economics}

\newcommand{\acemod}{\textsc{Ace}Mod\xspace}
\newcommand{\amtrac}{\textsc{AMT}ra\textsc{C-19}\xspace}

\newcolumntype{L}[1]{>{\raggedright\let\newline\\\arraybackslash\hspace{0pt}}m{#1}}
\newcolumntype{C}[1]{>{\centering\let\newline\\\arraybackslash\hspace{0pt}}m{#1}}
\newcolumntype{R}[1]{>{\raggedleft\let\newline\\\arraybackslash\hspace{0pt}}m{#1}}

\usepackage[affil-it]{authblk}

\def\arraystretch{1.3}
\begin{document}

\title{Modelling transmission and control of the COVID-19 pandemic in Australia}

\author[1]{ \ Sheryl L. Chang}
\author[1]{ \ Nathan Harding}
\author[1]{ \ Cameron Zachreson}
\author[1]{ \ Oliver M. Cliff}
\author[1,2,*]{ \ Mikhail Prokopenko}
\affil[1]{Centre for Complex Systems, Faculty of Engineering, \protect\\ University of Sydney, Sydney, NSW 2006, Australia}
\affil[2]{Marie Bashir Institute for Infectious Diseases and Biosecurity, \protect\\ University of Sydney, Westmead, NSW 2145, Australia}
\affil[*]{Corresponding author: mikhail.prokopenko@sydney.edu.au (ORCID: 0000-0002-4215-0344)}

\date{}

\maketitle

\begin{abstract}
There is a continuing debate on relative benefits of various mitigation and suppression strategies aimed to control the spread of COVID-19. Here we report the results of agent-based modelling using a fine-grained computational simulation of the ongoing COVID-19 pandemic in Australia. This model is calibrated to match key characteristics of COVID-19 transmission. An important calibration outcome is the age-dependent fraction of symptomatic cases, with this fraction for children found to be one-fifth of such fraction for adults. We apply the model to compare several intervention strategies, including restrictions on international air travel, case isolation, home quarantine, social distancing with varying levels of compliance, and school closures. School closures are not found to bring decisive benefits, unless coupled with high level of social distancing compliance. We report several trade-offs, and an important transition across the levels of social distancing compliance, in the range between 70\% and 80\% levels, with compliance at the 90\% level found to control the disease within 13--14 weeks, when coupled with effective case isolation and international travel restrictions.
\end{abstract}


\clearpage

\section*{Introduction}

The coronavirus disease 2019 (COVID-19) pandemic is an ongoing crisis caused by severe acute respiratory syndrome coronavirus 2 (SARS-CoV-2). The first outbreak was detected in December 2019 in Wuhan, the capital of Hubei province, rapidly followed by the rest of Hubei and all other provinces in China.  Within mainland China the epidemic was largely controlled by mid- to late March 2020, having generated more than 81,000 cases (cumulative incidence on 20 March 2020~\cite{NHC}). This was primarily due to intense quarantine and social distancing measures, including: isolation of detected cases; tracing and management of their close contacts; closures of potential zoonotic sources of SARS-CoV-2; strict traffic restrictions and quarantine on the level of entire provinces (including suspension of public transportation, closures of airports, railway stations, and highways within cities); cancellation of mass gathering activities; and other measures aimed to reduce transmission of the infection~\cite{wang2020novel,WHOChina,ChinaCDC}. 

Despite the unprecedented domestic control measures, COVID-19 was not completely contained and the disease reached other countries. On 31 January 2020, the epidemic was recognised by the World Health Organization (WHO) as a public health emergency of international concern, and on 11 March 2020 the WHO declared the outbreak a pandemic~\cite{WHO11032020}. 
Effects of the COVID-19 pandemic have quickly spilled over from the healthcare sector into international trade, tourism, travel, energy and finance sectors, causing {profound social and economic ramifications~\cite{Lenzen2020}}. While worldwide public health emergencies have been declared and mitigated in the past---e.g., the ``swine flu'' pandemic in 2009~\cite{Longini2005,Ferguson,nsoesie2012sensitivity,nsoesie2014systematic}---the scale of socio-economic disruptions caused by the unfolding COVID-19 pandemic is unparalleled in recent history. 

Australia began to experience most of these consequences, with the number of confirmed COVID-19 cases crossing 1,000 by 21 March 2020, whilst (at that time) doubling every three days, and the cumulative incidence growth rate averaging 0.20 per day during the first three weeks of March 2020 (Appendix~A in SI). 
In response, the Australian government introduced strict intervention measures in order to prevent the epidemic from continuing along such trends and to curb the devastating growth seen in other COVID-19 affected nations.
Nevertheless, there is an ongoing debate on the utility of specific interventions (e.g., school closures), the low compliance with social distancing measures (e.g., reduction of mass gatherings), and the optimal combination of particular health intervention options balanced against social and economic ramifications, and restrictions on civil liberties.  
In the context of this debate, there is an urgent requirement for rigorous and unbiased evaluations of available options. The present study makes a contribution towards this requirement and provides timely input into the Australian pandemic response discussion. Specifically, we develop a large-scale Agent-Based Model (ABM) capturing
salient features of COVID-19 transmission in Australia, and use it to evaluate the effectiveness of
non-pharmaceutical interventions with respect to the population's compliance with the suggested measures.

Governments around the world are presently fighting the spread of COVID-19 within their jurisdictions by developing, applying and adjusting multiple variations on pandemic intervention strategies.
While these strategies vary across nations, they share fundamental approaches which are adapted by national healthcare systems, aiming at a broad adoption within societies.  In the absence of a COVID-19 vaccine, as pointed out by Ferguson et al. \cite{Ferg2020}, mitigation policies may include case isolation of patients and home quarantine of their household members,  social distancing of the individuals within specific age groups (e.g., the elderly, defined as older than 75 years), as well as people with compromised immune systems or other vulnerable groups. In addition, suppression policies may require an extension of case isolation and home quarantine with social distancing of the entire population. Often, such social distancing is supplemented by school and university closures. 

Our primary objective is an evaluation of several intervention strategies that have been  deployed in Australia, or have been considered for a deployment:
restriction on international arrivals (``travel ban''); 
in-home case isolation (CI) of ill individuals; 
home quarantine (HQ) of family members of ill individuals;  
social distancing (SD) at various population compliance levels up to and including 100\%, a full lockdown; 
school closures (SC), which affect the behavior of school children as well as their parents and teachers. 
We explore these intervention strategies independently and in various combinations, as detailed in Methods. Each scenario is traced over time and compared to the baseline model in order to quantify its potential to curtail the epidemic in Australia. Our aims are to identify minimal effective levels of social distancing compliance, and to determine the potential impact of school closures on the effectiveness of intervention measures.

Stochastic agent-based models have been established as robust tools for tracing the fine-grained effect of heterogeneous intervention policies in diverse epidemic and pandemic settings~\cite{Halloran2002,eubank2004modelling,Ferguson,Longini2004,Longini2005,GermannKadauEtAl2006,barrett2010integrated,balcan2010modeling,chao2010flute}, including for policy advice currently in place in the USA and the UK~\cite{Ferg2020}. 
In this study, we follow the ABM approach to quantitatively evaluate and compare several mitigation and suppression measures, using a high-resolution individual-based computational model calibrated to key characteristics of COVID-19 pandemics.
The approach uses a modified and extended agent-based model, \acemod, previously developed and validated for simulations of pandemic influenza in Australia \cite{Cliff2018,Zachreson2018,Harding2020,Zachreson2020}.
The epidemiological component,  \amtrac, is developed and calibrated specifically to COVID-19 via reported invariants (outputs) such as the growth rate above.
Importantly, our sensitivity analysis shows that key epidemiological outputs from our model (e.g., the growth rate, $R_0$, generation time, etc.) are robust to uncertainty in the input parameters (e.g., the natural history of the disease, fraction of symptomatic cases, etc.).

In investigating possible effects of various intervention policies, we are able to provide clear and tangible goals for the population and government to pursue in order to mitigate the pandemic within Australia. The key result, based on a comparison of several intervention strategies, is an actionable transition across the levels of social distancing compliance, identified in the range between 70\% and 80\% levels.
A compliance of below 70\% is unlikely to succeed for any duration of social distancing, while a compliance at the 90\% level is found to control the disease within 13--14 weeks, when coupled with effective case isolation, home quarantine, and international travel restrictions.
We validate these results by a comparison with the actual epidemic and social distancing compliance observed in Australia. In doing so, we confirm that the model has successfully predicted the cumulative incidence as well as the timing of both the incidence and prevalence peaks.
Moreover, we illustrate trade-offs between these levels and duration of the interventions, and between the interventions' delay and their duration. Specifically, our simulations suggest that a three-day delay in introducing strict intervention measures lengthens their required duration by over three weeks on average, i.e., 23.56 days (with standard deviation of 11.167).

\section*{Results}

We present results of the high-resolution (individual-based) pandemic modelling in Australia, including a comparative analysis of intervention strategies. 
As discussed above, we performed our analysis using \acemod, an established Australian-census calibrated ABM that captures fine-grained demographics and social dynamics~\cite{Cliff2018,Zachreson2018,Harding2020,Zachreson2020}.
The epidemiological component of our model, \amtrac, was developed and calibrated to match key characteristics of COVID-19 (see Methods).

The input parameters were calibrated to generate key characteristics in line with reported epidemiological data on COVID-19.  
We primarily calibrated by comparing these epidemiological characteristics to the mean of output variables, inferred from Monte Carlo simulations during non-intervention periods, with confidence intervals (CIs) constructed by bootstrapping (i.e., random sampling with replacement) with the bias-corrected percentile method~\cite{tibshirani1993introduction}. 

The key output variables, inferred in concordance with available data, include: a reproductive number $R_0$ of $2.77$, 95\% CI [2.73, 2.83], $N = 6,315$; a generation period $T_{gen}$ of 7.62 days, 95\% CI [7.53, 7.70], $N = 6,315$; a growth rate of cumulative incidence during a period of sustained and unmitigated local transmission at $\dot{C} = 0.167$ per day, 95\% CI [0.164, 0.170], $N = 20$; and an attack rate in children of $A_c = 6.154\%$, 95\% CI [6.15\%, 6.16\%], $N = 20$. The relatively narrow confidence intervals reflect the intrinsic stochasticity of the simulations carried out for the default values of input parameters. The broad range of possible variations in response to changes in the input parameters, as well as the
robustness of the model and its outcomes, are established by  the sensitivity analysis (see Appendix~D in SI). This is followed by validation against actual epidemic timeline in Australia (see Appendix~H in SI), confirming that the adopted parametrization is acceptable.

\subsection*{Baseline}
A trace of the baseline model --- no interventions whatsoever --- is shown in Fig.~\ref{SD_comparison_tot}, with clear  epidemic peaks in both incidence (Fig.~\ref{SD_comparison_tot}.a) and prevalence (Fig.~\ref{SD_comparison_tot}.b) evident after 105--110 days from the onset of the disease in Australia, i.e., occurring around mid-May 2020.
The scale of the impact is very high, with nearly 50\% of the Australian population showing symptoms. This baseline scenario is provided only for comparison, in order to evaluate the impact of interventions, most of which were already in place in Australia during the early phase of epidemic growth. To re-iterate, we consider timely intervention scenarios applicable to the situation in Australia at the end of March 2020, with the number of confirmed COVID-19 cases crossing 2,000 on 24 March 2020, and the growth rate of cumulative incidence $\dot{C}$ {averaging 0.20 per day during the first three weeks of March}. We observe that the simulated baseline generates $\dot{C} \approx 0.17$ per day, in a good agreement with actual dynamics. 

\subsection*{Case isolation: CI, and home quarantine: HQ}
\begin{figure}[ht]
	\centering
  \includegraphics[clip, trim=8.1cm 0cm 7.8cm 2.0cm, width=1.0\textwidth]{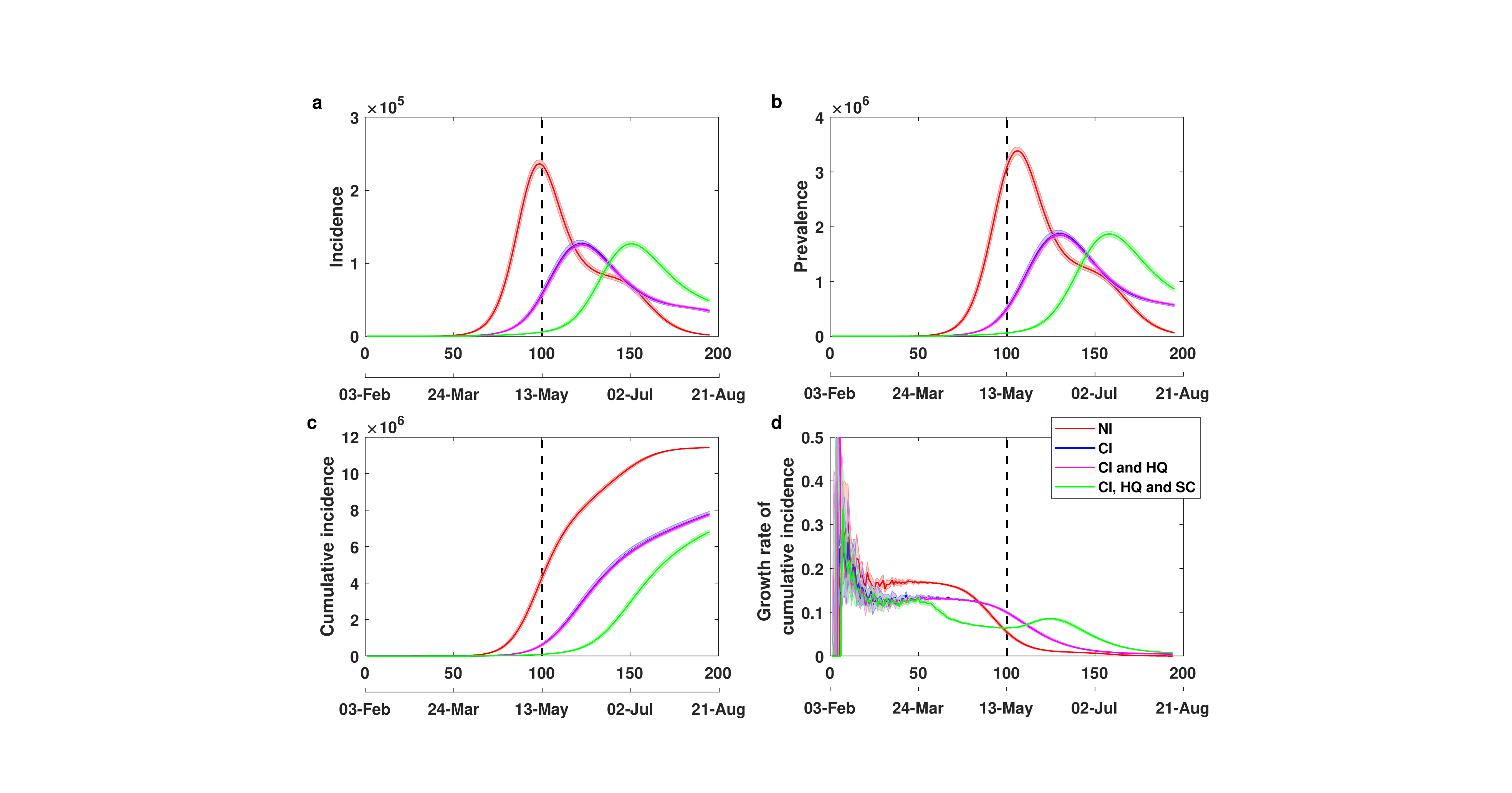}
\caption{\textbf{Effects of case isolation, home quarantine and school closures.} A combination of the case isolation (CI) and home quarantine (HQ) measures delays epidemic peaks and reduce their magnitude, in comparison to no interventions (NI), whereas school closures (SC) have short-term effect. Several baseline and intervention scenarios, traced for \textbf{a} incidence, \textbf{b} prevalence, \textbf{c} cumulative incidence, and \textbf{d} the daily growth rate of cumulative incidence $\dot{C}$, shown as average (solid) and 95\% confidence interval (shaded) profiles, over 20 runs. The 95\% confidence intervals are constructed from the bias corrected bootstrap distributions. The strategy with school closures combined with case isolation lasts 49 days (7 weeks), marked by a vertical dashed line. Restrictions on international arrivals are set to last until the end of each scenario. The alignment between simulated days and actual dates may slightly differ across separate runs.}
\label{SD_comparison_tot}
\end{figure}

All the following interventions include restrictions on international arrivals, triggered by the threshold of 2,000 cases. 
Three mitigation strategies are of immediate interest: 
\begin{enumerate}[label=(\roman*),noitemsep]
	\item case isolation,
	\item in-home quarantine of household contacts of confirmed cases,
	\item school closures, combined with (i) and (ii).
\end{enumerate}
These strategies are shown in Fig. \ref{SD_comparison_tot}, with the duration of the SC strategy set as 49 days (7 weeks), starting when the threshold of 2,000 cases is reached. The case isolation coupled with home quarantine delays the epidemic peak by about 26 days on average (e.g., shifting the incidence peak from day 97.5 to day 123.2, Fig.~\ref{SD_comparison_tot}.a, and the prevalence peak from day 105 to day 130.7, Fig.~\ref{SD_comparison_tot}.b, on average). In addition, CI combined with HQ reduces the height of the epidemic peak by around 47--49\%. The main contributing factor is case isolation, as adding home quarantine, with 50\% in-home compliance, to case isolation of 70\% symptomatic individuals, delays the epidemic peak by less than three days on average. The overall attack rate resulting from the coupled policy is also reduced in comparison to the baseline scenario (Fig.~\ref{SD_comparison_tot}.c). However, case isolation and home quarantine, even when coupled together, are not effective for epidemic suppression, with prevalence still peaking in millions of symptomatic cases (1.873M), Fig.~\ref{SD_comparison_tot}.b. Such an outcome would have completely overburdened the Australian healthcare system~\cite{moss2020modelling}. 

\subsection*{School closures: SC}
Adding school closures to the case isolation and home quarantine approach also does not achieve a significant reduction in the overall attack rate (Fig. \ref{SD_comparison_tot}). The peaks of both incidence (Fig.~\ref{SD_comparison_tot}.a) and prevalence (Fig.~\ref{SD_comparison_tot}.b) are delayed by about four weeks (about 27 days for both incidence and prevalence). However, their magnitudes remain practically the same, due to a slower growth rate of cumulative incidence (Fig. \ref{SD_comparison_tot}.d). This is observed irrespective of the commitment of parents to stay home (Appendix~G in SI).  We also traced the dynamics resulting from the SC strategy for two specific age groups: children and individuals over 65 years old, shown in Appendix~G in SI.  The four-week delays in occurrence of the peaks are observed across both age groups, suggesting that there is a strong concurrence in the disease spread across these age groups. We also observe that under the SC strategy coupled with case isolation and home quarantine, the magnitude of the incidence peak for children increases by about 7\%  shown in Appendix~G in SI (Supplementary Fig.~9.a). This may be explained by increased interactions of children in household and community social mixing environments, when schools are closed. Under this strategy, there is no difference in the magnitude of the incidence peak for the older age group (Appendix~G in SI, Supplementary Fig.~10.a). We also note that the considered interventions succeed in reducing a relatively high variance in the incidence fraction of symptomatic older adults, thus, reducing the epidemic potential to adversely affect this age group specifically.

In short, the only tangible benefit of school closures, coupled with case isolation and home quarantine, is in delaying the epidemic peak by four weeks, at the expense of a slight increase in the contribution of children to the incidence peak.
While school closures are considered an important part of pandemic influenza response, our results suggest that this strategy is much less effective in the context of COVID-19. The gains are further reduced by other societal costs of school closures, e.g., drawing their parents employed in healthcare and other critical infrastructure away from work.
There is, nevertheless, one more possible benefit of school closures, discussed in the context of the population-wide social distancing in Appendix~G in SI.
 
\subsection*{Social distancing: SD}
\begin{figure}[ht]
	\centering
  \includegraphics[clip, trim=8.1cm 0cm 7.8cm 2.0cm, width=1.0\textwidth]{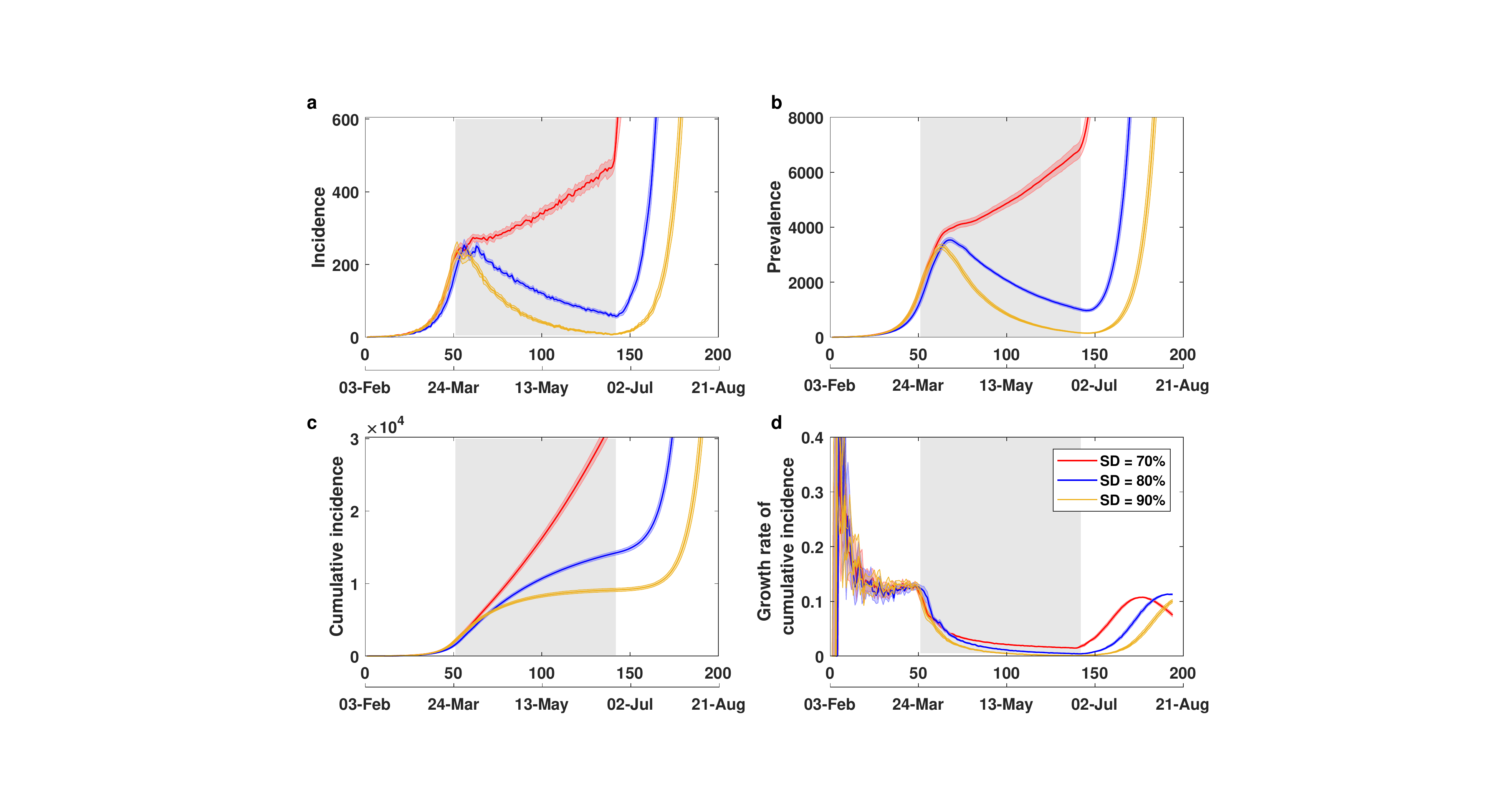}
\caption{\textbf{Effects of social distancing.} Strong compliance with social distancing (at 80\% and above) effectively controls the disease during the suppression period, while lower levels of compliance (at 70\% or less) do not succeed for any duration of the suppression. A comparison of social distancing strategies, coupled with case isolation, home quarantine and international travel restrictions, across different compliance levels (70\%, 80\% and 90\%). Duration of each social distancing (SD) strategy is set to 91 days (13 weeks), shown as a grey shaded area between days 51 and 142; (the start and end days of SD varied across stochastic runs: for 70\% SD the last day of suppression was 141.4 on average; for 80\% SD it was 144.2; and for 90\% SD it was 141.5, see Source Data file). Case isolation, home quarantine and restrictions on international arrivals are set to last until the end of each scenario. Traces include \textbf{a} incidence, \textbf{b} prevalence, \textbf{c} cumulative incidence, and \textbf{d} the daily growth rate of cumulative incidence $\dot{C}$, shown as average (solid) and 95\% confidence interval (shaded) profiles, over 20 runs. The 95\% confidence intervals are constructed from the bias corrected bootstrap distributions. The alignment between simulated days and actual dates may slightly differ across separate runs.}
\label{SD_compliance}
\end{figure}

Next, we examine the effects of population-wide social distancing in combination with case isolation and restrictions on international arrivals. 
Here, we present the effects of different compliance levels on the epidemic dynamics.
Low compliance levels, set at less than 70\%, did not show any potential to suppress the disease in the considered time horizon (28 weeks), while the total lockdown, that is, complete social distancing at 100\%, managed to reduce the incidence and prevalence to zero, after 49 days of the mitigation. However, 
 because it is unrealistic to expect 100\% compliance in the Australian context,
we focus on the practically achievable compliance levels: 70\%, 80\% and 90\%, with their duration set to 91 days (13 weeks), shown in Fig.~\ref{SD_compliance}. 

Importantly, during the time period that the SD level is maintained at 70\%, the disease is not controlled, with the numbers of new infected cases (incidence) remaining in hundreds, and the number of active cases (prevalence) remaining in thousands. Thus, 70\% compliance is inadequate for reducing the effective reproductive number below $1.0$. In contrast, the two higher levels of SD, 80\% and 90\%, are more effective at suppressing both prevalence and incidence during the 13-week social distancing period.   

Figure~\ref{SD_compliance} contrasts these three levels of SD compliance, ``zooming in'' into the key time period, immediately following the introduction of social distancing.  Crucially, there is a qualitative difference between the lower levels of SD compliance (70\%, or less), and the  higher levels (80\%, or more). For the SD compliance set at 80\% and 90\%, we observe a reduction in both incidence~(Fig.~\ref{SD_compliance}.a) and prevalence~(Fig.~\ref{SD_compliance}.b), lasting for the duration of the strategy (91 days). 
With SD compliance of 80\% the disease is not completely eliminated, but incidence is reduced to less than 100 new cases per day, with prevalence below 1,000 by the end of the suppression period, Fig.~\ref{SD_compliance}.b. It is important to note that while the disease is suppressed during the period over which social distancing is in effect, resurgence of transmission is likely unless complete or near-complete elimination has been achieved upon cessation of social distancing measures. Our results suggest that this level of compliance would succeed in eliminating the disease in Australia if the strategy was implemented for a longer period, e.g., another 4--6 weeks.  

The 90\% SD compliance practically controls the disease, bringing both incidence and prevalence to very low numbers of isolated cases (and reducing the effective reproductive number to nearly zero). It is possible for the epidemic to spring back to significant levels even under this level of compliance, as the remaining sporadic cases indicate a potential for endemic conditions. We do not quantify these subsequent waves, as they develop beyond the immediately relevant time horizon. Nevertheless, we do share the concerns expressed by the Imperial College COVID-19 Response Team: ``The more successful a strategy is at temporary suppression, the larger the later epidemic is predicted to be in the absence of vaccination, due to lesser build-up of herd immunity'' \cite{Ferg2020}.  Given that the herd immunity threshold is determined by $1 - 1/R_0$~\cite{anderson1985vaccination}, the extent required to build up collective immunity for COVID-19, assuming $R_0 = 2.77$, may be estimated as 0.64, that is, 64\% of the population becoming infected or eventually immunised. 

The cumulative incidence for the best achievable scenario (90\% SD compliance coupled with case isolation, home quarantine, and restrictions on international arrivals) settles in the range of 8,000 -- 10,000 cases during the suppression period, with resurgence still possible at some point after intervention measures are relaxed, Fig.~\ref{SD_compliance}.c. The range of cumulative incidence at the end of the suppression is 8,313 -- 10,090 over 20 runs, with the mean of 9,122 cases and 95\% CI [8,898, 9,354], constructed from the bias corrected bootstrap distribution (see Source Data file). In terms of case numbers, this is an outcome several orders of magnitude better than the worst case scenario, developing in the absence of the combined mitigation and suppression strategies. 

We compare two sets of scenarios. In our primary scenarios, aligned with the actual epidemic curves in Australia, the social distancing measures are triggered by 2,000 confirmed cases. In alternative scenarios, the strict suppression measures are initiated earlier, being triggered by crossing the threshold of 1,000 cases (Appendix~H.1 in SI). The best agreement between the actual and simulation timelines is found to match a delayed but high ($90\%$) SD compliance, appearing to be followed from 24 March 2020, after a three-day period with a weaker compliance which commenced on 21 March 2020 when the international travel restrictions were introduced, as shown in Fig.~\ref{actual} and detailed in Appendix~H.2 in SI.  
For the 1,000 case threshold scenario, we present the effects of different SD compliance levels (70\% and 90\%) on the spatial distribution of cases on day 60.  These are shown in Appendix~I in SI, as choropleth maps of the four largest Australian Capital Cities: Sydney, Melbourne, Brisbane and Perth.

\begin{figure}[ht]
\vspace*{-1cm}
\centering
    \includegraphics[clip, trim=7.7cm 0cm 7.8cm 0cm, width=1.0\textwidth]{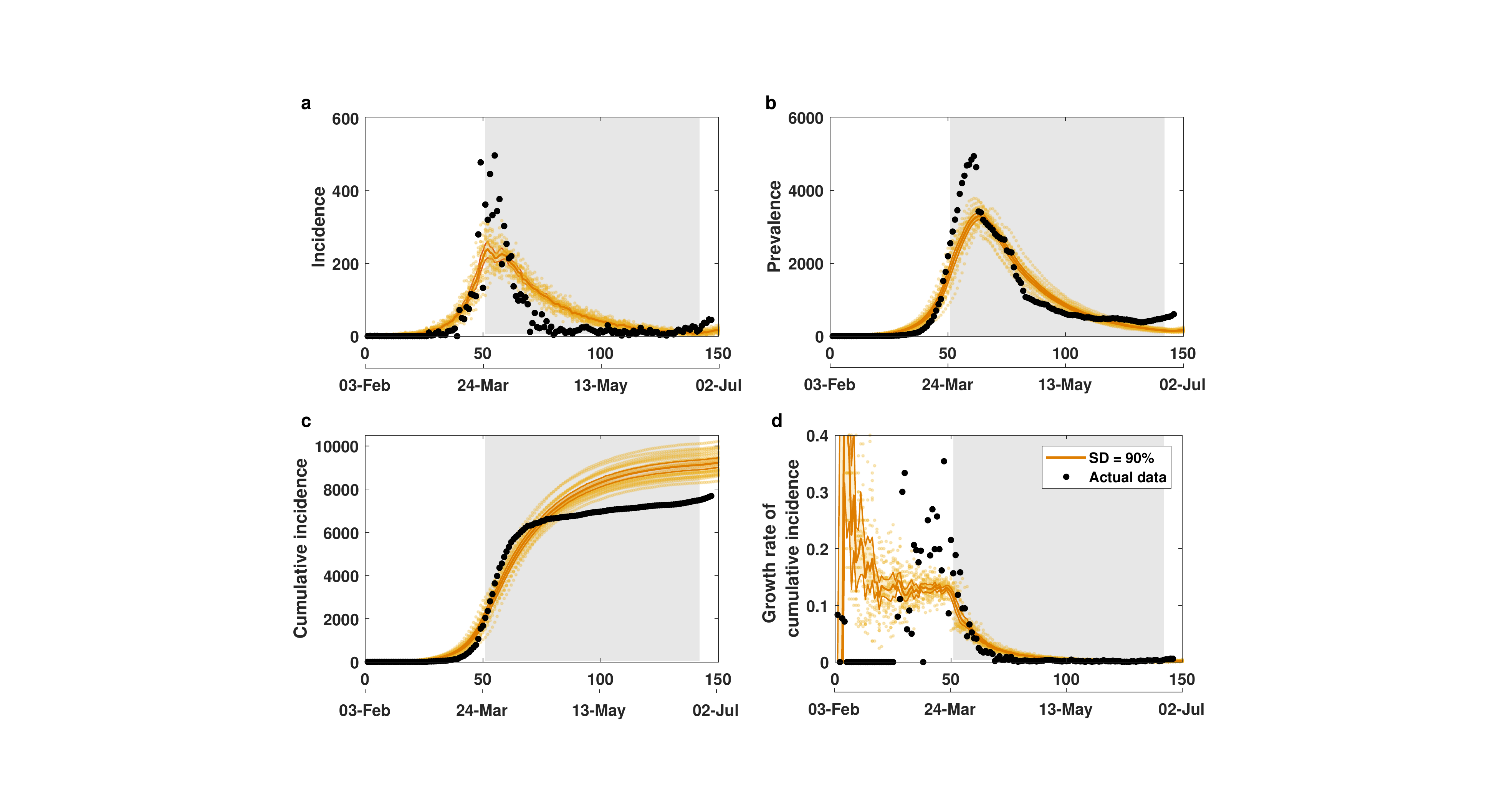}
\caption{\textbf{Model validation with actual data.} A comparison between actual epidemic curves in Australia (black dots, shown until 28 June 2020), and the primary simulation scenario, using a threshold of 2,000 cases (crossed on 24 March 2020) and following 90\% of social distancing (SD), coupled with case isolation, home quarantine, and international travel restrictions, shown until early July 2020 (yellow colour). Duration of the SD strategy is set to 91 days (13 weeks), shown as a grey shaded area. Case isolation, home quarantine, and restrictions on international arrivals are set to last until the end of the scenario. Traces include \textbf{a} incidence, \textbf{b} prevalence, \textbf{c} cumulative incidence, and \textbf{d} daily growth rate of cumulative incidence, shown as average (solid), 95\% confidence interval (thin solid) profiles, as well as the ensemble of 20 runs (scatter). The 95\% confidence intervals are constructed from the bias corrected bootstrap distributions. The alignment between simulated days and actual dates may slightly differ across separate runs. Data sources:~\cite{wiki-merged,hopkins}.}
    \label{actual}
	\vspace*{-1cm}
\end{figure}

It is clear that there is a trade-off between the level of SD compliance and the duration of the SD strategy: the higher the compliance, the more quickly incidence is suppressed.  
Both 80\% and 90\% compliance levels control the spread within reasonable time periods: 18-19 and 13-14 weeks respectively. 
In contrast, lower levels of compliance (at 70\% or less) do not succeed for any duration of the imposed social distancing limits. 
This quantitative difference is of major policy setting importance, indicating a sharp transition in the performance of these strategies in the region between 70\% and 80\%. 
 
\begin{figure}[ht]
	\centering
  \includegraphics[clip, trim=1.6cm 0cm 1.0cm 0cm, width=1.0\textwidth]{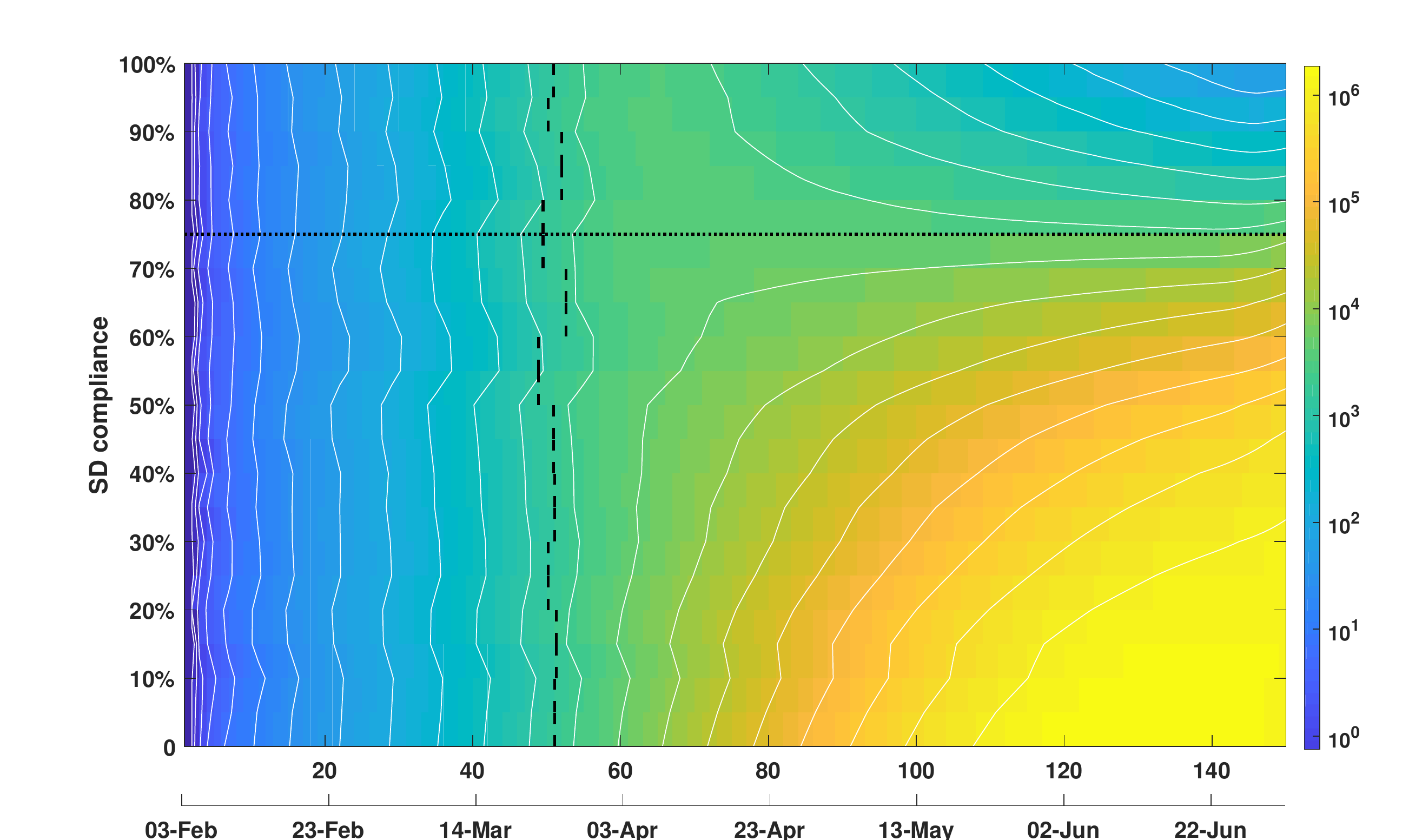}
\caption{\textbf{Phase transition across the levels of social distancing compliance.} Colour image plot of disease prevalence as a function of time (horizontal axis) and social distancing (SD) compliance (vertical axis). A phase transition is observed between 70\% and 80\% SD compliance (marked by a dotted line). For SD compliance levels below 80\%, the prevalence continues to grow after social distancing is implemented, while for compliance levels at or above 80\% the prevalence declines, following a peak formed after approximately two months. The colours correspond to log-prevalence, traced from the epidemic's onset until the end of the suppression period.  The isolines trace contours with constant values of log-prevalence. Vertical dashes mark the time when threshold of 2,000 is crossed, triggering SD, averaged over 20 runs for each SD level.  Social distancing is coupled with case isolation, home quarantine and international travel restrictions. The alignment between simulated days and actual dates may slightly differ across separate runs.}
\label{ptrans}
\end{figure}

Referring to Fig.~\ref{ptrans}, the identified transition across the levels of compliance with social distancing may also be interpreted as a tipping point or a phase transition~\cite{yeomans1992statistical}. 
Various critical phenomena have been discovered previously in the context of epidemic models, often interpreting epidemic diffusion in statistical-mechanical terms, for example, as percolation within a network~\cite{newman1999scaling,newman2002spread,Harding2018,Harding2020b}.
The transition across the levels of SD compliance is similar to percolation transition in a forest-fire model with immune trees~\cite{Guisoni2011}.  
Distinct epidemic phases are evident in Fig.~\ref{ptrans} at a certain percolation threshold between the
SD compliance of 70\% and 80\%, at which the critical regime exhibits the effective reproductive number $R_{\text{eff}} = 1.0$. That is, crossing this regime signifies moving into the phase where the epidemic is controlled, i.e., reducing $R_{\text{eff}}$ below $1.0$.

We do not attempt to establish a more precise level of required compliance between 70\% and 80\%. Such a precision would be of lesser practical relevance than the identification of 80\% {compliance} as the minimal acceptable level of social distancing, with 90\% providing a shorter timeframe. The robustness of these results is established by sensitivity analysis presented in Appendix~D.2 in SI.

In addition, a three-day delay in introducing strong social distancing measures is projected to extend the required suppression period by approximately three weeks, beyond the 91-day period considered in the primary scenario (see Appendix~H in SI). Finally, we report fractions of symptomatic cases across mixing contexts (Appendix~J in SI), with the infections through households being predominant. Notably, the household fractions steadily increase with the strengthening of SD compliance, while the corresponding fractions of infections in the workplace and school environments decrease.

\subsection*{Summary}

In short, the best intervention approach identified in our study is a combination of international travel restrictions, case isolation, home quarantine, and social distancing with at least 80\%--90\% compliance for a duration of approximately 91 days (13 weeks). These measures have been implemented in Australia to a reasonable degree, however, it is unclear if testing throughput and contact tracing resources are sufficient to facilitate effective interventions if incidence increases substantially. For these reasons, it is our conclusion that social distancing is likely to continue to be the instrumental line of defense against COVID-19 in Australia. In our study, compliance levels below 80\% resulted in higher prevalence at the end of suppression period, and increasing incidence during the social distancing period.

We point out that our results are relevant only for the duration of the mitigation and suppression, and a resurgence of the disease is possible once these interventions cease, as shown in Fig.~\ref{SD_compliance}. We also note that a rebound in the incidence and prevalence post-suppression period is not unavoidable: more efficient and large-scale testing methods are expected to be developed in several months, and so the resultant contact tracing and case isolation are likely to prevent a resurgence of the disease. The international travel restrictions are assumed to stay in place. Hence, we do not quantify the precise impact of control measures beyond the selected time horizon (28 weeks), 
aiming to provide immediately relevant insights. Furthermore, our results should not be seen as policies optimised over all possible parameter combinations, but rather as a clear demonstration of the extent of social distancing required to reduce incidence and prevalence over two to six months.

\section*{Discussion}

In this study we simulated several possible scenarios of COVID-19 pandemic's spread in Australia. The model,  \amtrac, was calibrated to known pandemic dynamics, and accounted for age-dependent attack rates, a range of reproductive numbers, age-stratified and social context dependent transmission rates, household clusters and other social mixing contexts, symptomatic-asymptomatic distinction, 
and other relevant epidemiological parameters.
An important calibration result was the need for age-dependent fractions of symptomatic agents, with the fraction of symptomatic children found to be one-fifth of that of the adults.

We reported several findings relevant to COVID-19 mitigation and suppression policy setting.  The first implication is that the effectiveness of school closures is limited (under our assumptions on the age-dependent symptomatic fractions and the infectivity in children), producing a four-week delay in epidemic peak, without a significant impact on the magnitude of the peak, in terms of incidence or prevalence. The temporal benefit of this  delay may be offset not only by logistical complications, but also by some increases in the fractions of both children and older adults during the period around the incidence peak. As the clinical picture of COVID-19 in children continues to be refined~\cite{hoang2020covid}, these findings may benefit from a re-evaluation when more extensive pediatric data become available.

The second implication is related to the social distancing (SD) strategy, which showed little benefit for lower levels of compliance (at 70\% or less) --- these levels do not  produce epidemic suppression for any duration of the social distancing restrictions. Only when the SD compliance levels exceed 80\%, there is a reduction in incidence and prevalence. Our modelling results indicate existence of an actionable transition across these strategies in the range between 70\% and 80\%. In other words, increasing a compliance level just by 10\%, from 70\% to 80\%, may effectively control the spread of COVID-19 in Australia, by reducing the effective reproductive number to near zero (during the suppression period). 

We also reported a trade-off between the compliance levels and the duration of SD mitigation, with 90\% compliance significantly reducing incidence and prevalence after a shorter period of 91 days (13 weeks). Although a resurgence of the disease is possible once these interventions cease, we believe that this study could facilitate a timely planning of effective intervention  and exit strategies. In particular, 
this study contributed to the report, ``Roadmap to Recovery'', presented to the Australian Federal Government on 29 April 2020, providing evidence for a comparison between two options. Rather than recommending ``a single dominant option for pandemic response in Australia'', the roadmap pointed out considerable and evolving uncertainties, and presented two strategies: (i) a state by state elimination of local community transmissions (with the restrictions remaining for a longer duration, but achieving lower cases and greater public confidence), and (ii) controlled adaptation aimed at some minimal level of symptomatic cases within the health system capacity (with phased and adaptive lifting of restrictions, beginning as early as May 15th 2020, but acknowledging the high likelihood of prolonged global circulation of SARS-CoV-2)~\cite{Roadmap}. However, a precise evaluation of detailed exit strategies, as well as the probability of elimination, lies outside of the scope of our study.

Future research will address several limitations of our study, including a more fine-grained implementation of natural history of the disease, reducing uncertainty around the transmissibility and infectivity in young people, incorporation of more recent Australian Bureau of Statistics (ABS) data from 2020, and an account of hospitalisations and in-hospital transmissions. We also hope to trace specific spatial pathways and patterns of epidemics, in order to enable a detailed understanding of how the infection spreads in diverse circumstances and localities, with the aim to identify the best ways to locate and curtail the pandemic spread in Australia.  It would be interesting to contrast our ABM with network-based approaches: while both frameworks depart from the compartmental fully mixed models in capturing specific interactions affecting the infection spread, there are differences in describing the context dependence and ways to intervene~\cite{newman2002spread,cauchemez2011role}. In network-based models, the most effective interventions have been found to be those which reduce the diversity of interactions~\cite{meyers2003applying}, and can be modelled by changes in the topology of contact networks~\cite{small2020modelling}. And so one future direction would be a comparison of the epidemic and intervention thresholds across the ABM and network-based models.
Other avenues lead to analysis of precursors and critical thresholds for possible emergence of new strains, as well as various ``change points'' in the spreading rate~\cite{antia2003role,Harding2018,dehning2020inferring}, studies of genomic surveillance data interpreted as complex networks~\cite{epjb,salmonella,rockett2020revealing}, dynamic models of social behaviour in times of health crises~\cite{mossong2008social,chang2020game,gros2020strategies}, and investigations of global socioeconomic effects of the COVID-19 pandemic~\cite{Ferg2020b,Dignum,Lenzen2020}.

\clearpage

\section*{Methods}

\acemod, the Australian Census-based Epidemic Model, employs a discrete-time and stochastic agent-based model to investigate complex outbreak scenarios across the nation over time.
The \acemod simulator comprises over 24 million software agents, each with attributes of an anonymous individual
(e.g., age, gender, occupation, susceptibility and immunity to diseases), as well as contact rates within different social contexts (households, household clusters, local neighbourhoods, schools, classrooms, workplaces). The set of generated agents captures average characteristics of the real population, e.g., \acemod is calibrated to the Australian Census data (2016) with respect to key demographic statistics. In addition, the \acemod simulator has integrated layered school attendance data from the Australian Curriculum, Assessment and Reporting Authority (ACARA), within a realistic and dynamic interaction model, comprising both mobility and human contacts. These social mixing layers represent the demographics of Australia as close as possible to the Australian Bureau of Statistics (ABS) and other datasets, as described in Appendix~F in SI.

Potential interactions between spatially distributed agents are represented using data on mobility in terms of commuting patterns (work, study and other activities), adjusted to increase precision and fidelity of commute networks \cite{Fair2019}.  Each simulation scenario runs in 12-hour cycles (``day'' and ``night'') over the 196 days (28 weeks) of an epidemic, and agents interact across distinct social mixing groups depending on the cycle, for example, in working groups and/or classrooms during a ``day'' cycle, and their households, household clusters, and local communities during the ``night'' cycle. The interactions result in transmission of the disease from infectious to susceptible individuals: given the contact and transmission rates, the simulation computes and updates agents' states over time, starting from initial infections, seeded in international airports around Australia \cite{Cliff2018,Zachreson2018}. The simulation is implemented in C++11, using the g++ compiler (GCC) 4.9.3 and GNU Autotools (autoconf 2.69, automake 1.15), running under CentOS release 6.9 (upstream Red Hat 4.4.7-18) on a High Performance Computing (HPC) service and utilising 4264 cores of computing capacity. Post-processing of simulation results is carried out with MATLAB R2020a.

Simulating disease transmission in \acemod requires both (i) specifics of local transmission dynamics, dependent on individual health characteristics of the agents, such as susceptibility and immunity to disease, driven by their transmission and contact rates across different social contexts; and (ii) a natural disease history model for COVID-19, i.e., the infectivity profile from the { exposure}, to the peak of infectivity, and then to recovery, for a single symptomatic or asymptomatic infected individual. The infectivity of agents is set to exponentially rise and peak at 5 days, after { two days of zero infectivity}.  The symptoms are set to last up to $12$ days post the infectivity peak, during which time infectiousness linearly decreases to zero. The probability of transmission for asymptomatic/presymptomatic agents is set as $0.3$ of that of symptomatic individuals; and the age-dependent fractions of symptomatic cases are set as $\sigma_c = 0.134$ for children, and $\sigma_a = 0.669$ for adults.  These parameters were calibrated to available estimates of key transmission characteristics of COVID-19 spread,  implemented in \amtrac, the Agent-based Model of Transmission and Control of the COVID-19 pandemic in Australia. 

\subsection*{Calibration}

Despite several similarities with influenza, COVID-19 has a number of notable differences, specifically in relation to transmissions across children, its reproductive number $R_0$, incubation and generation periods, proportion of symptomatic to asymptomatic cases,
the infectivity of the asymptomatic and presymptomatic individuals, etc. (see Appendix~B in SI). While uncertainty around the reproductive number $R_0$, the incubation and generation periods, as well as the age-dependent attack rates of the disease, have been somewhat reduced~\cite{WHOChina,ChinaCDC,guan2020clinical}, there is still an ongoing effort in estimating the extent to which people without symptoms, or exhibiting only mild symptoms, might contribute to the spread of the coronavirus~\cite{Lieabb3221}. Furthermore, the question whether the ratio of symptomatic to total cases is constant across age groups, especially children, has not been explored in studies to date, remaining another critical unknown.  
 
Thus, our first technical objective was to calibrate the \amtrac model for specifics of COVID-19 pandemic, in order to determine key disease transmission parameters of  \amtrac, so that the resultant dynamics concur with known estimates. In particular, we investigated a range of the reproductive number $R_0$ (the number of secondary cases arising from a typical primary case early in the epidemic). The range $[2.0, 2.5]$ has been initially reported by the WHO-China Joint Mission on Coronavirus Disease 2019~\cite{WHOChina}. Several studies estimated that before travel restrictions were introduced in Wuhan on 23 January 2020, the median daily reproduction number $R_0$ in Wuhan was $2.35$, with 95\% confidence interval (CI) of [1.15, 4.77]~\cite{kucharski2020early}.  On 15 April 2020, Australian health authorities reported $R_0$ in the range [2.6, 2.7]~\cite{Roadmap}, while more recent Australian and international studies investigated $R_0$ in the range [2.5, 3.5]~\cite{moss2020modelling,Roadmap,gros2020strategies,dehning2020inferring}. For example, a median $R_0 = 3.4$ (CI [2.4, 4.7]) was used in a model of the COVID-19 spread in Germany~\cite{dehning2020inferring}, while the estimates reviewed by Liu et al.~\cite{taaa021} ranged from 1.4 to 6.49, with a mean of 3.28 and a median of 2.79. In our model, $R_0$, our output variable $y_1$, was investigated in the range between 1.94 and 3.12, see Table~\ref{tab:kappa-R0}, by varying a scaling factor $\kappa$ responsible for setting the contagiousness of the simulated epidemic, as explained in Appendix~C in SI~\cite{Cliff2018,Harding2020}.

We aimed for the generation period $T_{gen}$, i.e., our output variable $y_2$, to stay in the range $[6.0, 10.0]$~\cite{li2020early,huang2020epidemic,Ferg2020}. This is {also} in line with the reported mean serial interval of 7.5 days (with 95\% CI of 5.3 to 19)~\cite{li2020early}. 

In addition, we aimed to keep the resultant daily growth rate of cumulative incidence $\dot{C}$, output variable $y_3$, around $0.2$ per day, in order to be consistent with the disease dynamics reported in Australia and internationally (see Appendix~A in SI). Our focus was to characterise the rate of a rapid infection increase during the sustained but unmitigated local transmission. This calibration target was chosen at the time, mid-March 2020, to complement $R_0$ and the generation period, given the lack of data on the epidemic peak values, and fragmented patient recovery and prevalence data. By that time, despite different initial conditions and disease surveillance regimes, as well as diversity of case definitions, several countries exhibited a similar growth pattern. This suggested that a steady growth rate of approximately $0.2$ per day may provide a consistent calibration target during the early growth period, with seven out of the top eight affected nations settling around this rate after a noisy transient (except South Korea where the initial growth had the cluster nature, following a superspreading event~\cite{shim2020transmission}).

Another key constraint was a low attack rate in children, $A_c$, i.e., our output variable $y_4$, reported to be in single digits. For example, only $2.4$\% of all reported cases in China were children, while a study in Japan observed that ``it is remarkable that there are very few child cases aged from 0--19 years'', with only $3.4\%$ of all cases in this age group~\cite{mizumoto2020age}.  

The calibration was aimed at satisfying our key constraints, given by the expected ranges of output variables. In doing so, we varied several ``free'' parameters, such as transmission and contact rates, the fraction of symptomatic cases (making it age-dependent), the probability of transmission for both symptomatic and asymptomatic agents, and the infectivity profile from the exposure. 
 Specifically, we explored the time to infectivity peak, our input parameter $x_1$, in proximity to known estimates of the mean incubation period, i.e., between 4 and 7 days, calibrating the time to peak to $5.0$ days. In several studies, the mean incubation period was reported as 5.2 days, 95\% CI, 4.1 to 7.0 \cite{li2020early}, while being  distributed around a mean of approximately 5 days within the range of 2--14 days with 95\% CI \cite{linton2020incubation}.  We also varied the symptoms' duration after the peak of infectivity, i.e., recovery period, our input parameter $x_2$, between 7 and 21 days, and calibrated it at $12.0$ days, on a linearly decreasing profile from the peak.
 
The contact and transmission rates across various mixing contexts 
detailed in Appendices~C and~E in SI. The probability of transmission for asymptomatic/presymptomatic agents, our input parameter $x_3$,  was set as 0.3 of that of symptomatic individuals (lower than in the \acemod influenza model), having been explored between 0.05 and 0.45. Both symptomatic and asymptomatic  infectivity profiles were changed to increase exponentially after a latent period of two days, reaching the infectivity peak after five days, with the onset of symptoms distributed across agents during this period, see Appendix~C in SI.  

The fraction of symptomatic cases, our input parameter $x_4$, was investigated in the range between 0.5 and 0.8, and set to two-thirds of the total cases ($\sigma_a = 0.669$), which concurs with several studies. For example, the initial data on 565 Japanese citizens evacuated from Wuhan, China, who were symptom-screened and tested, indicated that 41.6\% were asymptomatic, with a lower bound estimated as 33.3\% (95\% CI: 8.3\%, 58.3\%)~\cite{nishiura2020estimation}. The proportion of asymptomatic cases on the Diamond Princess cruise ship was estimated in the range between 17.9\% (95\% credible interval (CrI): 15.5--20.2\%) to 39.9\%
(95\% CrI: 35.7--44.1\%)~\cite{mizumoto2020estimating}, noting that most of the passengers were 60 years and older, and more likely to experience more symptoms. The modelling study of Ferguson et al.~\cite{Ferg2020} also set the  fraction of symptomatic cases to $\sigma = 0.669$. 

However, we found that our output variables were within the expected ranges only when this fraction is age-dependent, with the fraction of symptomatic cases among children, our input parameter $x_5$,  calibrated to one-fifth of the one for adults, that is, $\sigma_c = 0.134$ for children, and $\sigma_a = 0.669$ for adults. This calibration outcome per se,  achieved after exploring the range $\sigma_c \in [0.05, 0.25]$, is in agreement with the reported low symptomaticity in children worldwide, and the observation that ``children are at similar risk of infection as the general population, though less likely to have severe symptoms''~\cite{bi2020epidemiology}. Another study of epidemiological characteristics of  2,143 pediatric patients in China noted that over 90\% of patients were asymptomatic, mild, or moderate cases~\cite{dong2020epidemiological}. 

\begin{figure}
	\centering 					 
 \includegraphics[clip, trim=8.1cm 0cm 7.5cm 0cm, width=1.0\textwidth]{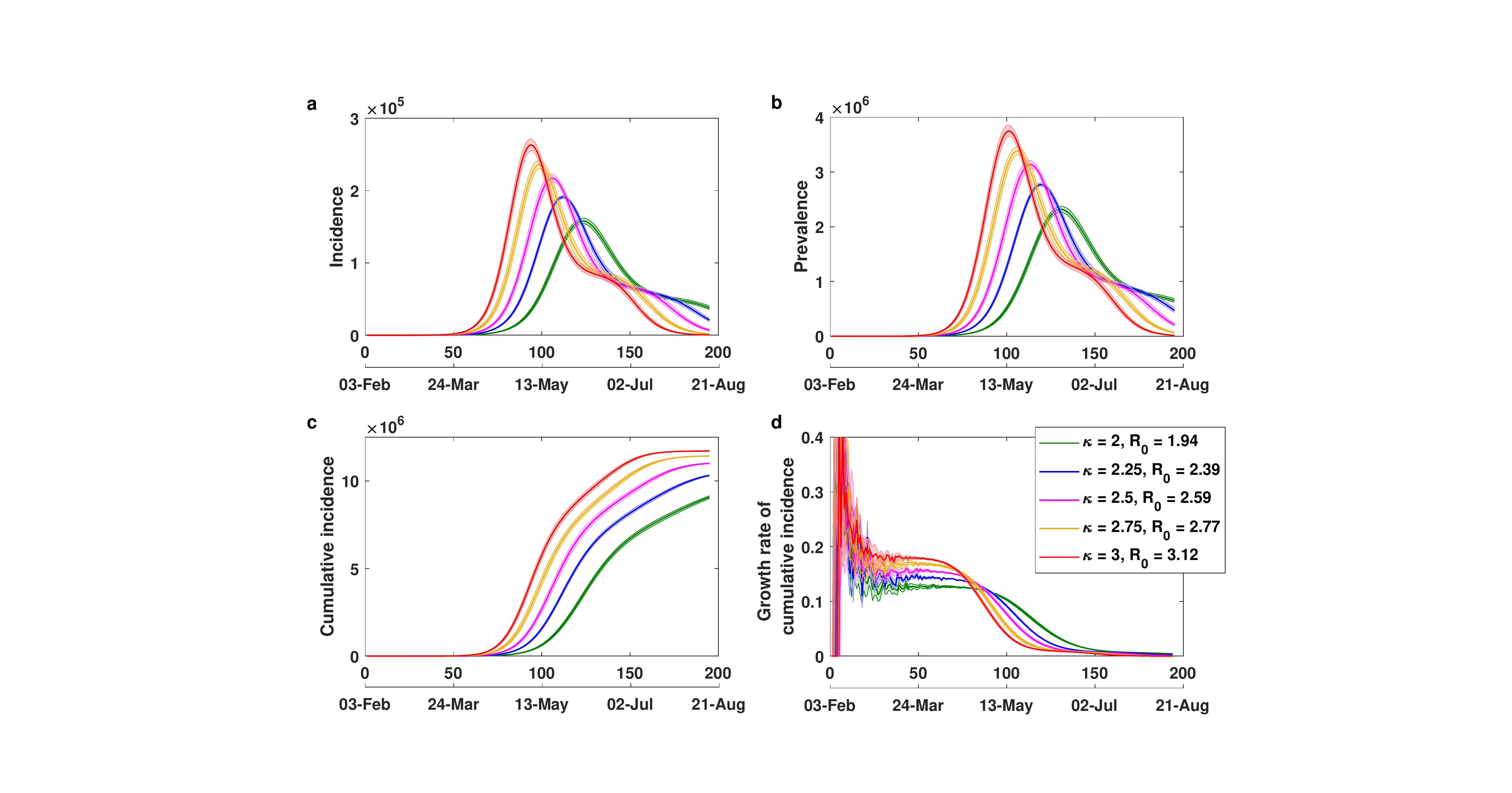}
\caption{\textbf{Model calibration with scaling factor $\kappa$.} Tracing \textbf{d} the expected growth rate of cumulative incidence $\dot{C}$ per day, while varying scaling factor $\kappa$ (proportional to the reproductive number $R_0$), with \textbf{a}~incidence, \textbf{b} prevalence, and \textbf{c} cumulative incidence.  Averages over 20 runs are shown as solid profiles, with 95\% confidence intervals shown as shaded profiles. The 95\% confidence intervals are constructed from the bias corrected bootstrap distributions. The alignment between simulated days and actual dates may slightly differ across separate runs.}
\label{calibrated_tot}
\end{figure}

\begin{figure}
	\centering
\includegraphics[clip, trim=7.7cm 0cm 7.5cm 0cm, width=1.0\textwidth]{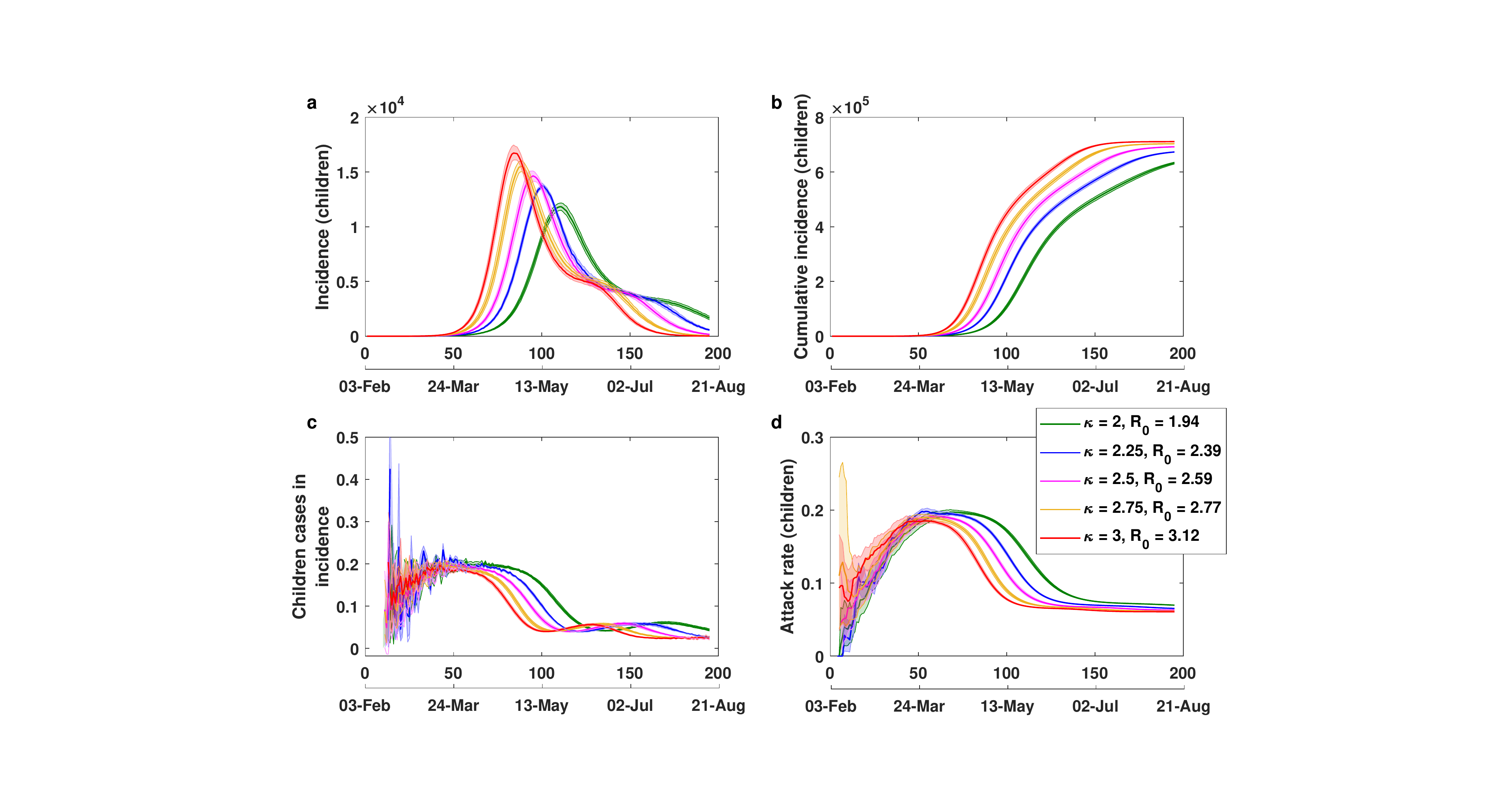}
\caption{\textbf{Model calibration: epidemic curves for children.} Tracing \textbf{d} the attack rate in children, while varying scaling factor $\kappa$ (i.e., reproductive number $R_0$), with \textbf{a} incidence, \textbf{b} cumulative incidence, and \textbf{c} incidence fraction for children.   Averages over 20 runs are shown as solid profiles, with 95\% confidence intervals shown as shaded profiles.  The 95\% confidence intervals are constructed from the bias corrected bootstrap distributions. The alignment between simulated days and actual dates may slightly differ across separate runs.}
\label{calibrated_cld}
\end{figure}

In summary, this combination of parameters resulted in the dynamics that matched several COVID-19 pandemic characteristics. It produced the following estimates and their confidence intervals, constructed from the bias corrected bootstrap distribution:
\begin{enumerate}[noitemsep]
\item the reproductive number $R_0 = 2.77$, with 95\% CI [2.73, 2.83] (sample size $N = 6,315$); 
\item the generation period $T_{gen} = 7.62$ days, with 95\% CI [7.53, 7.70] ($N = 6,315$); 
\item the growth rate of cumulative incidence, determined at day 50, during a period of sustained unmitigated local transmission, $\dot{C} = 0.167$ per day, with 95\% CI [0.164, 0.170] and range [0.156, 0.182] ($N = 20$);  
\item the attack rate in children $A_c = 6.154\%$, with 95\% CI [6.15\%, 6.16\%] and range [6.14\%, 6.16\%] ($N = 20$). 
\end{enumerate}
Both the reproductive number and the generation period correspond to $\kappa = 2.75$ (see Table~\ref{tab:kappa-R0} for other values of $\kappa$).
The resultant dynamics are shown in Figures~\ref{calibrated_tot} and~\ref{calibrated_cld}. The sensitivity analysis of the output variables to changes in the input parameters is presented in Appendix~D.1 in SI. We point out that, in hindsight, one may choose more comprehensive calibration targets and refine the model with different parametrizations. The model presented in this study was calibrated by 24 March 2020, using Australian and international incidence and prevalence data from two preceding months, as well as constraints on the output variables detailed above. At the time, a limited testing capacity resulting in possible under-reporting of cases (especially pediatric) may have introduced a potential bias in model calibration. Nevertheless, the study is described here as an approach which succeeded in accurately predicting the epidemic peaks in Australia in early April (both incidence and prevalence), while providing timely advice on relevant pandemic interventions.

\bgroup
\begin{table}
	\caption{{ The reproductive number $R_0$ and the generation period $T_{gen}$ (with 95\% confidence intervals, constructed from the bias corrected bootstrap distribution), for various values of the scaling parameter $\kappa$.}}
	\label{tab:kappa-R0}
	\vspace{1mm}
	\centering
\resizebox{0.8\textwidth}{!}{%
	{\raggedright
	 \noindent
	{ 
	 \begin{tabular}{lcccccc}
	$\kappa$ & $R_0$ & 95\% CI & & $T_{gen}$ & 95\% CI & sample size \\
	\hline 
	2.00 & 1.94 & [1.91, 1.98] & & 6.92 & [6.81, 7.02] & 6,274 \\
	2.25 & 2.39 & [2.35, 2.44] & & 7.36 & [7.27, 7.45] & 6,372 \\
	2.50 & 2.59 & [2.54, 2.64] & & 7.46 & [7.37, 7.55] & 6,351 \\
  2.75 & 2.77 & [2.73, 2.83] & & 7.62 & [7.53, 7.70] & 6,315 \\
	3.00 & 3.12 & [3.10, 3.21] & & 7.74 & [7.66, 7.82] & 6,413 \\
\hline
	\end{tabular}
	} 
	}
}
\end{table}
\egroup

\subsection*{Fraction of local community transmissions}

We trace scenarios of COVID-19 pandemic spread in Australia, initiated by passenger arrivals via air traffic from overseas. This process maintains a stream of new infections at each time step, set in proportion to the average daily number of incoming passengers at that airport \cite{Zachreson2018,Harding2020}. These infections occur probabilistically, generated by binomial distribution $B(P,N)$, where $P$ and $N$ are selected to generate one new infection within a 50 km radius of the airport, per $0.04\%$ of incoming arrivals on average. 

In a separate study~\cite{rockett2020revealing}, we directly compared the fractions of local transmissions detected by our ABM with the genomic sequencing of SARS-CoV-2, carried out in a subpopulation of infected patients within New South Wales (NSW), the most populous state of Australia, until 28 March 2020. 
Only a quarter of sequenced cases was deemed to be locally acquired (cases who had not travelled overseas in the 14 days before illness onset), and this was in concordance with the trace obtained from our ABM model. 
Specifically, having simulated the five-week period preceding intervention measures, we inferred all local transmission links within households (HH), household clusters (HC), and local government areas which map to the census statistical areas (SA). 
Each directed link connecting two infected individuals in the same mixing context is detected if the infected agents share the same HH, HC or SA identifier, and the direction is inferred using the relevant simulation time steps. 
Then the fraction of local community transmissions is determined as the ratio between the number of the inferred transmission links and the number of total infections during the corresponding time period. 
These fractions ranged between 18.6\% (std. dev. 2.9\%) for HH and HC combined, and 34.9\% (std. dev. 8.2\%) for all transmissions within HH, HC and SA, broadly agreeing with the fraction identified through genomic surveillance: 25.8\% for all local transmissions~\cite{rockett2020revealing}.

\subsection*{Sensitivity analysis}	

We performed our sensitivity analysis using the local (point-based) sensitivity analysis (LSA)~\cite{cacuci2003sensitivity}, as well as global sensitivity analysis with the Morris method (the elementary effect method)~\cite{morris1991factorial}. Each method computes the response of an ``output'' variable of interest, e.g., the generation period, to the change in an ``input'' parameter, e.g., the fraction of symptomatic cases. The response $F_{i,j}$  of the state variable $y_j$ to parameter $x_i$ from a scaled vector of all $k$ input parameters, $\mathbf{X} = [0,1]^k$, is determined as a finite difference 
\begin{equation}
F_{i,j} = \frac{y_j(x_1, x_2, \ldots, x_i + \Delta, x_{i+1}, \ldots, x_k) - y_j(\mathbf{X})}{\Delta}
\end{equation}
where $\Delta$ is a discretisation step, dividing each dimension of the parameter space. The distribution of each response $F_{i,j}$ is obtained by repeated random sampling with a number of simulation runs per step. In LSA, an input parameter is varied, while keeping other inputs set at their base points, i.e., default values. In the Morris method, an input parameter is varied at a number of different points sampled across the domains of other parameters. The mean $\mu^*_{i,j}$ of the absolute response $|F_{i,j}|$ serves to capture the influence of the parameter $x_i$ on the output $y_j$: a large mean suggests  a higher sensitivity. The standard deviation $\sigma_{i,j}$ of the response $F_{i,j}$ is a complementary measure of sensitivity: a large deviation indicates that the dependency between the input and output is nonlinear. In the Morris method, a large deviation may also indicate that the input parameter interacts with other parameters~\cite{wu2013sensitivity}. Importantly, the responses are not directly comparable across the output variables, and instead are ranked across the inputs for each output. A model is generally considered robust if most of the dependencies are characterised by low means and deviations, with the variations contained within acceptable ranges of the output variables. 
Appendix~D in SI summarises the investigated ranges and results of the sensitivity analysis.

\subsection*{Intervention strategies}

\subsubsection*{International travel restrictions}
In our model, restriction on international arrivals is set to be enforced from the moment when the number of confirmed infections exceeds the threshold of 2,000 cases. 
This concurs well with the actual epidemic timeline in Australia, which imposed a ban on all arrivals of non-residents, non-Australian citizens, from 9pm of 20 March 2020, with a requirement for strict self-isolation of returning citizens. The number of COVID-19 cases crossed 1,000 cases on 21 March 2020, and doubled to slightly over 2,000 on 24 March 2020, so the 2,000 threshold chosen on our model reflects a delay in implementing the measures.   
The restriction on international arrivals is included in modelling of all other strategies, and is not traced independently, as this mitigation approach is not under debate. 

\subsubsection*{Case isolation}
The case isolation { mitigation strategy assumes that}
70\% of symptomatic cases stay at home, reduce their non-household contacts by 75\% (so that their transmission rates decrease to 25\% of the {baseline} rate), and maintain their household contacts (i.e., their transmission rates within household remain unchanged). { The assumption that even relatively mild symptomatic cases are identified and isolated is justified by the practice adopted in Australia where a comprehensive disease surveillance regime was consistently implemented. This included screening of syndromic fever and cough in combination with exhaustive case identification and management, thus enabling early detection (e.g., more than 1\% of the Australian population has been tested for the coronavirus by early April 2020, and the numbers of tests conducted in Australia per new confirmed case of COVID-19, as well as per capita, remain among the highest in the world)~\cite{lokuge2020exit,moss2020modelling,rockett2020revealing,hopkins}.}

\subsubsection*{Home quarantine}
In our model of the home quarantine strategy for household contacts of index cases 
we allow compliance to vary within affected households (i.e., at the individual level). 
{In our implementation, 50\% of individuals will comply with home quarantine if a member of their household becomes ill. We simulate this as a reduction to 25\% of their usual non-household contact rates, and a consequent doubling of their contact rates within the household.}
Both case isolation and home quarantine are assumed to be in force from the first day of the epidemic, as has been the case in Australia.

\subsubsection*{Social distancing}
If an individual complies with social distancing, all working group contacts are removed, and all non-household contact rates are set to 50\% of the baseline value, while keeping contact rates within households unaltered. To simulate imposition of the intervention policy by the federal government,
the SD strategy is triggered by crossing the threshold of 2,000 cases (matching the actual timeline on 24 March 2020). An alternative threshold of 1,000 cases, matching the actual numbers reported on 21 March 2020, is considered to evaluate a delayed introduction of strong social distancing measures (Appendix~H in SI). In our study, we vary the SD compliance level from 0\% to 100\% (full lockdown), the compliance level is simply the percentage of individuals who comply with the measure.

\subsubsection*{School closures}
School closure removes students, their teachers, and a fraction of parents from daytime interactions (their corresponding transmission rates are set to zero), but increases their interaction rates within households (with a 50\% increase in household contact rates). All students and teachers are affected. For each affected household, a randomly selected parent chooses to stay at home, with a varying degree of commitment. Specifically, we compared 25\% or 50\% commitment, as in Australia there is no legal age for leaving school-age children home alone for a reasonable time, in relevant circumstances. 
This parameter range is concordant with the report of The Australian Bureau of Statistics (ABS), summarising a survey of household impacts of COVID-19 during early April 2020: the proportion of adults keeping their children home from school or childcare reached 24.9\%~\cite{ABS4940-1}. The upper considered limit, a half of parents, accounts for reasonable scenarios ensuring adequate parental supervision.
School closures are assumed to be followed with 100\% compliance, and may be concurrent with all other strategies described above. 
The SC strategy is also triggered by crossing the threshold of 2,000 cases.
We note that the Australian Federal Government has, so far, not enforced schools closures, and so we investigate the SC intervention separately from, or coupled with, the SD strategy. Hence, the evaluation of school closures provides an input to policy setting, rather than forecasts possible epidemic dynamics.

\subsubsection*{Compliance}
The agents affected by various compliance choices are determined in the beginning of each simulation run, with dependency between voluntary measures that does not allow an individual to be compliant with home quarantine if they are not also compliant with case isolation. Then the relevant changes in contact behaviour are applied to the selected agents in every 12-hour cycle. The restrictions are applied in a specific order: CI, HQ, SD, SC, with only the most relevant distancing assigned during each simulation cycle. 
For example, if a student is ill and in case isolation, the contact reduction factors associated with home quarantine, social distancing, and school closure would not apply to them, even if they are considered compliant with those measures.
The micro- and macro-distancing parameters defining the levels of compliance, together with the affected non-household and household contacts are summarised in Table~\ref{tab:micro-macro}.

\subsubsection*{Duration of measures}
While the case isolation and home quarantine strategies are assumed to last during the full course of the epidemic, we vary the duration of SD and/or SC strategies across a range of intervals, with a specific focus on 49 and 91 days, that is, 7 or 13 weeks.

\bgroup 
\begin{table}
	\caption{The micro- and macro-distancing parameters: macro-compliance levels and context-dependent micro-distancing levels.}
	\label{tab:micro-macro}
	\vspace{1mm}
	\centering
\resizebox{1.0\textwidth}{!}{%
	{\raggedright
	 \noindent
	{
	 \begin{tabular}{lcccc}
	 & Macro-distancing & \multicolumn{3}{c}{Micro-distancing contacts}  \\
	Strategy & Compliance levels & Household & Community & Workplace/School  \\
	\hline
	No intervention & 100\% & 100\% & 100\% & 100\% \\
	Case isolation & 70\% & 100\% & 25\% & 25\%  \\
  Home quarantine & 50\% & 200\% & 25\% & 25\%  \\
	School closure (children) & 100\% & 150\% & 150\% & 0\%  \\
	School closure (parents) & 25\% or 50\%& 150\% & 150\% & 0\%  \\
	Social distancing & 0--100\% & 100\% & 50\% & 0\% \\
	\hline 
	\end{tabular}
	} 
	}
}
\end{table}
\egroup

\section*{Acknowledgments}
The Authors are grateful to Stuart Kauffman, Edward Holmes, Joel C Miller, Paul Ormerod, Kristopher Fair, Philippa Pattison, Mahendra Piraveenan, Manoj Gambhir, Joseph Lizier, Peter Wang, John Parslow, Jonathan Nolan, Neil Davey, Vitali Sintchenko, Tania Sorrell,  Ben Marais, and Stephen Leeder, for  discussions of various intricacies involved in agent-based modelling of infectious diseases, and computational epidemiology in general. The Authors were supported through the Australian Research Council grants DP160102742 (SC, NH, OC, CZ, MP) and DP200103005 (MP). \acemod is registered under The University of Sydney's invention disclosure CDIP Ref. 2019-123.  \amtrac is registered under The University of Sydney's invention disclosure CDIP Ref. 2020-018. We are thankful for a support provided by High Performance Computing (HPC) service (Artemis) at the University of Sydney. This is a preprint of an article published in \emph{Nature Communications}. The final authenticated version is available online at: \url{https://www.nature.com/articles/s41467-020-19393-6}.

\section*{Author contributions}
SLC, NH, OC and MP developed and calibrated COVID-19 epidemiological model. CZ implemented intervention strategies. SLC carried out computational simulation, prepared figures, source and supplementary data files. SLC, CZ, OC and MP performed sensitivity analysis and tested the model. MP conceived the study and drafted the manuscript, with all authors contributing. All authors contributed to analysis and interpretation of the results, and gave final approval for publication.


%

\clearpage

\renewcommand{\figurename}{Supplementary Figure}
\renewcommand{\tablename}{Supplementary Table}

\section*{Supplementary Information}

\appendix

\section{COVID-19 pandemic in top 8 affected countries and Australia}
\label{intern} 
As of 21 March 2020, when significant intervention measures were introduced in Australia, over 285,000 cases have been confirmed worldwide, causing more than 11,500 deaths; and in a month, by 23 April, the total number has grown to exceed 2.628 million cases, with more than 183,400 deaths~\cite{dong2020interactive,hopkins}. By 21 March 2020, the disease established a sustained local transmission in many countries around the globe, with more than 180 countries and territories affected, including Italy, Spain, Iran, the United States, Germany, France, and South Korea as the top eight affected nations~\cite{dong2020interactive,hopkins}. 

\begin{figure}
	\centering
  \includegraphics[clip, trim=1.5cm 0cm 1.8cm 6cm, width=1.0\textwidth]{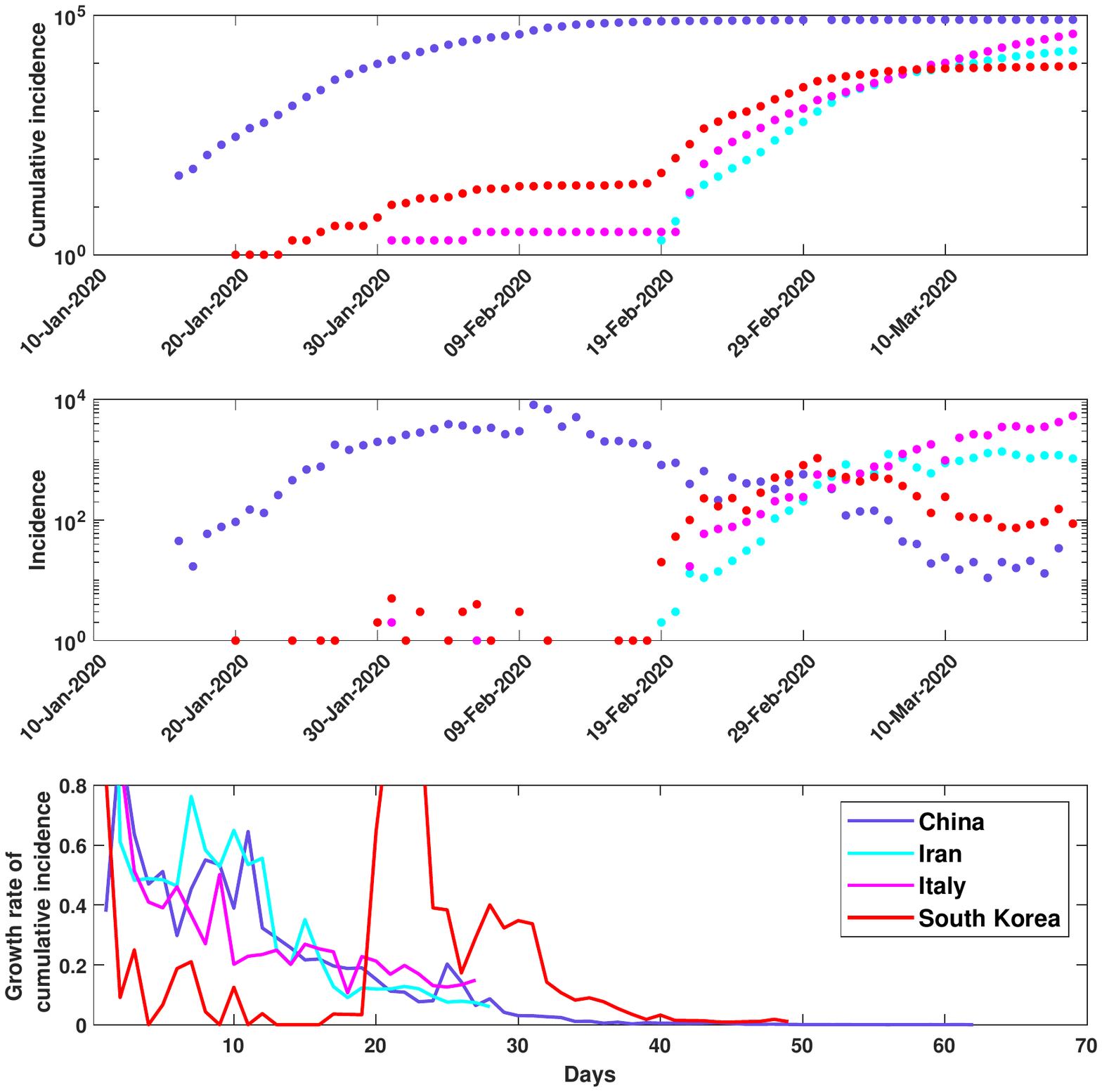}
\caption{\textbf{Early epidemic curves: China, Iran, Italy, South Korea.} Cumulative incidence (log scale), incidence (log scale), and daily growth rate of cumulative incidence (up to 19 March 2020). Days: since the day when the total number of confirmed cases exceeded five. Data sources:~\cite{wiki-merged}.}
\label{countries1}
\vspace*{-5cm}
\end{figure}

\begin{figure}
	\centering
  \includegraphics[clip, trim=1.5cm 0cm 1.8cm 11cm, width=1.0\textwidth]{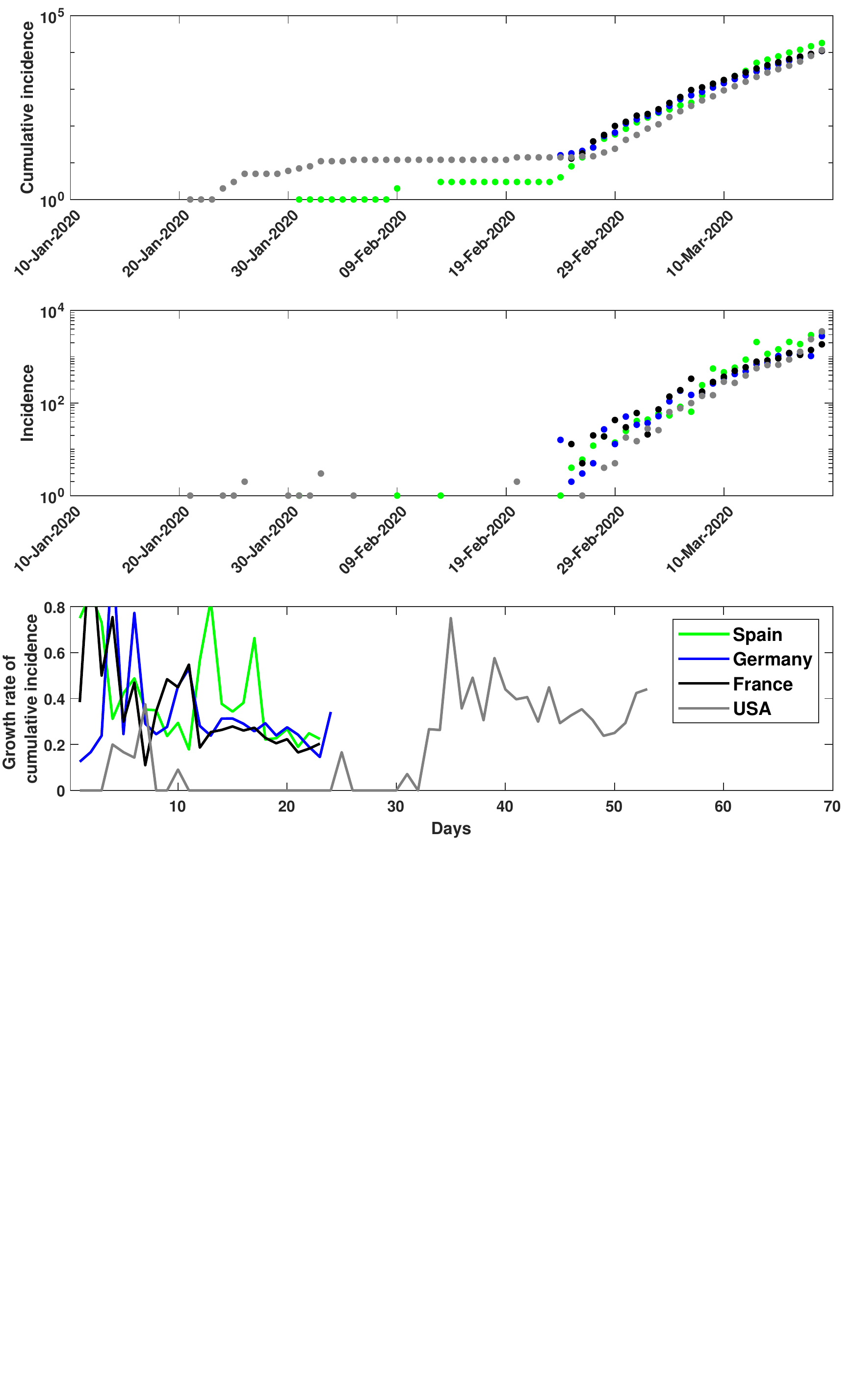}
\caption{\textbf{Early epidemic curves: Spain, Germany, France, USA.} Cumulative incidence (log scale), incidence (log scale), and {daily} growth rate of cumulative incidence (up to 19 March 2020). Days: since the day when the total number of confirmed cases exceeded five. Data sources:~\cite{wiki-merged}.}
\label{countries2}
\vspace*{-9cm}
\end{figure}

The scale of the COVID-19 pandemic has grown several orders of magnitude in a matter of weeks, from hundreds to thousands to tens of thousands, with the rate of these transitions varying across countries. Of particular interest to our study is the time periods when the epidemics are sustained locally in these countries, but before the effects of adopted intervention strategies are fully felt. One immediate observation is that during this period, the growth rate of cumulative incidence in many of the traced national epidemics is averaging within the range between 0.2 and 0.3 {per day}, that is, there are 20\% to 30\% daily increases in new cases on average. This is particularly evident for Spain, France, and Germany (Supplementary Fig.~\ref{countries2}), as well as China, Iran and Italy (Supplementary Fig.~\ref{countries1}). These average estimates provide approximate ``invariants'' and reduce uncertainty around key epidemiological parameters, required to calibrate disease transmission models, before investigating possible effects of various intervention policies. 

\begin{figure}
	\centering
  \includegraphics[clip, trim=1.2cm 0cm 1.9cm 5cm, width=1.0\textwidth]{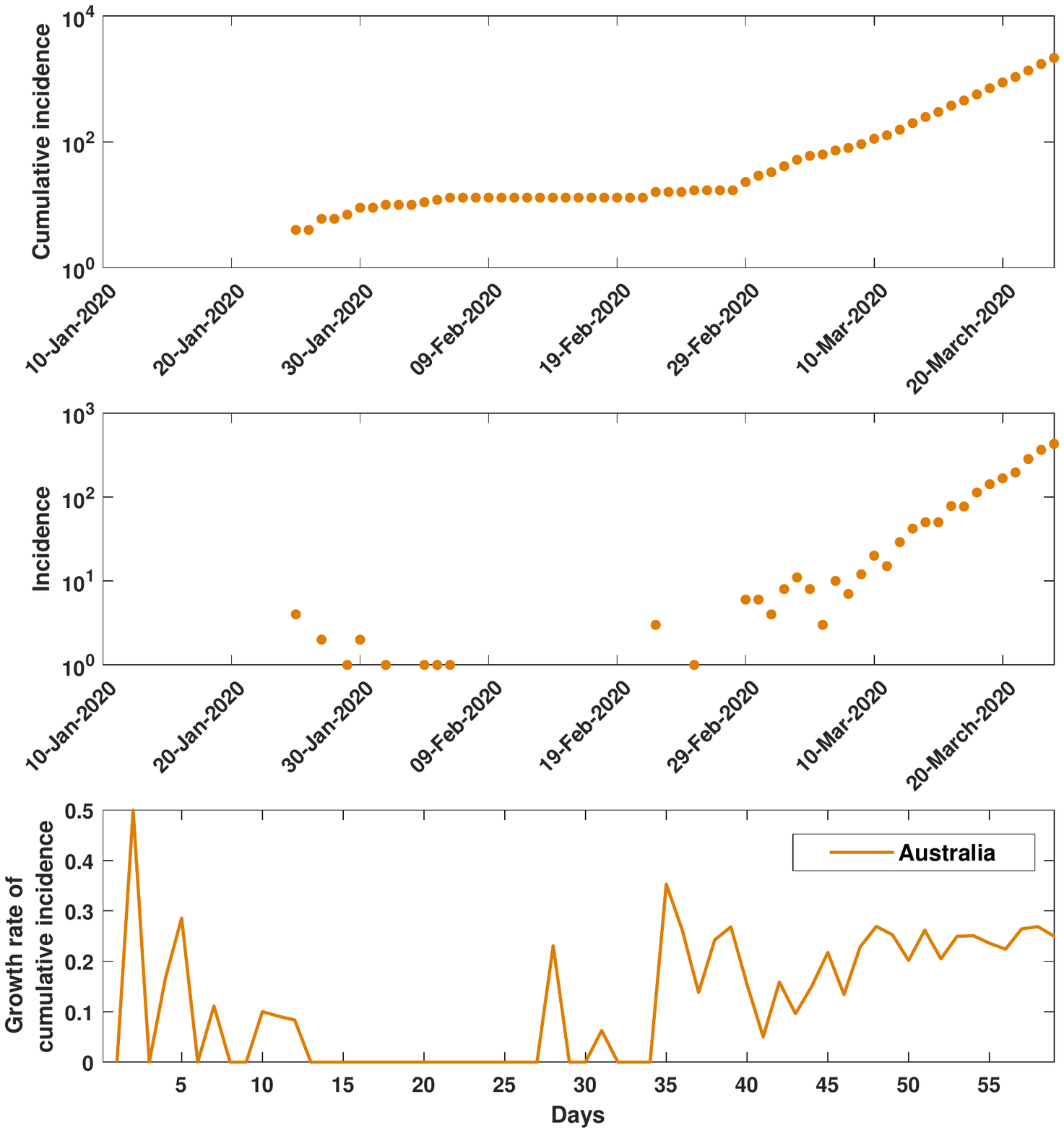}
\caption{\textbf{Early epidemic curves: Australia.} Cumulative incidence (log scale), incidence (log scale), and {daily} growth rate of cumulative incidence (up to 24 March 2020). Days: since the day when the total number of confirmed cases exceeded five. Data sources:~\cite{wiki-merged}.}
\label{countries3}
\vspace*{-5cm}
\end{figure}

Supplementary Figures~\ref{countries1} and \ref{countries2} trace cumulative incidence $C$, incidence, and {daily} growth rate of cumulative incidence $\dot{C} = [C(n+1) - C(n)]/C(n)$, for time step $n$, for the top eight affected countries (as of 21 March 2020): China, Iran, Italy, South Korea (Supplementary Fig.~\ref{countries1}), Spain, Germany, France, USA (Supplementary Fig.~\ref{countries2}). The time series begin from the day when the total number of confirmed cases exceeds five. Supplementary Figure~\ref{countries3} traces these time series for Australia. We reiterate that the fraction of imported cases in the overall transmission has been fairly high in Australia, dominating the community transmission, and so we paid particular attention to the daily growth rate in countries where the disease was also introduced predominantly through the air travel (i.e., down-weighting the rates in China and South Korea).

\section{Natural history of disease}
\label{hist}
The natural history of disease is a description of the disease {progression over time from exposure to recovery, in a single individual and in the absence of treatment}. In the past, the \acemod simulator has been used to model pandemic influenza within Australia, and here we detail modifications of  the natural history aimed to account for COVID-19 specifics, captured in \amtrac. We define several agent states: \textsc{susceptible}, \textsc{latent}, infectious \textsc{symptomatic}, infectious \textsc{asymptomatic}, and \textsc{recovered}. Consequently, the natural history model considers three distinct phases. The first phase is the latent period during which individuals are infected but unable to infect others, set in the COVID-19 model as two days. The second phase is the period characterised by an exponentially increasing infectivity, from 0\% to 100\% over three days (see Supplementary Fig.~\ref{fig:nhd}).
 The day on which an individual becomes ill is chosen probabilistically: 30\% of agents will change their state to symptomatic one day after exposure, 50\% after two days, and the remaining 20\% will become symptomatic on day three. In order to reflect the presence of mildly symptomatic cases, the model does not follow a canonical definition of an incubation period (the period between exposure to an infection and the appearance of the first symptoms). Instead, it distributes the onset of symptoms across the agents, and increases their infectivity to a peak over a number of days. The time to peak, five days, is chosen to align with an empirically estimated average incubation period, while mild symptoms may be detectable even before the peak. Upon reaching its peak, the infectivity decreases linearly over 12 more days (third phase), until the recovery, with immunity, occurs after 17 days. Finally, we assume that asymptomatic cases are 30\% as infectious as symptomatic cases. Unlike influenza, where we assume that the asymptomatic fraction is the same for adults as for children, for the SARS-COV-2 coronavirus we assume that while 67\% of adult cases are symptomatic, a significantly lower fraction (13.4\%) is symptomatic in children.

\begin{figure}[ht]
    \centering
    \includegraphics[clip, trim=2.3cm 0cm 2.3cm 0cm, width = 1.0\linewidth]{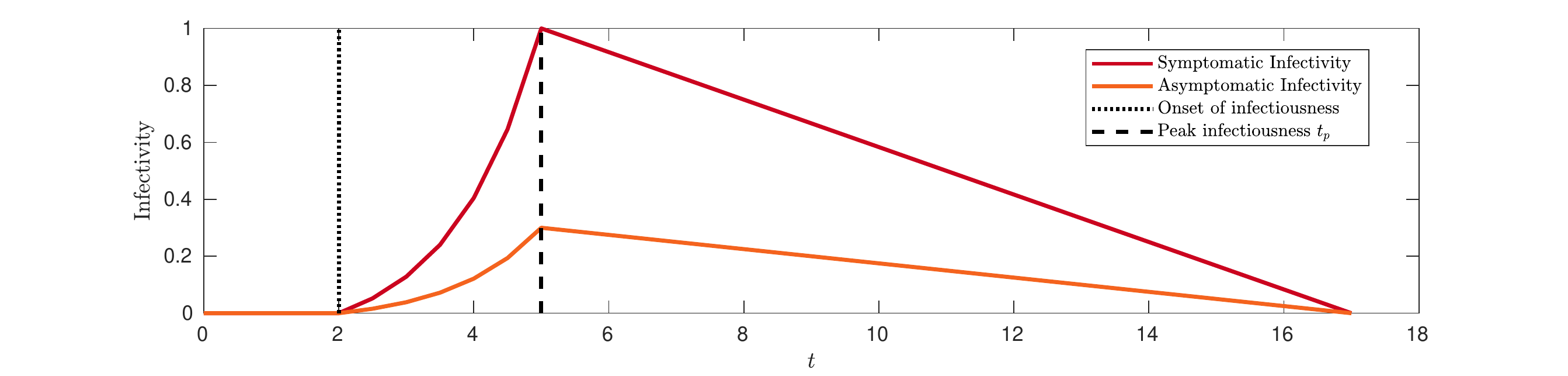}
    \caption{\textbf{Model of the natural history of COVID-19.} Profile of the infectivity, for both symptomatic and asymptomatic cases. After two days, individuals become infectious, with the infectivity rising exponentially until its peak at five days. After this peak, the infectivity linearly decreases, with full recovery occurring at 17 days. At comparable points within the natural history of disease, asymptomatic individuals are 30\% as infectious as symptomatic individuals.}
    \label{fig:nhd}
\end{figure}

\vspace*{-2mm}
\section{Transmission model and reproductive number}
\label{trans}
The primary dynamics of { \amtrac} are the infection transmissions. At each time-step the simulator determines the probability of infection for an individual, based on the infection levels in each of their mixing contexts. At each time step we consider all daytime or all nighttime contexts. Let $X_i(n)$ be a random variable describing the state of individual $i$ at time step $n$. At each time step we calculate $p_i(n) = P(X_i(n) = \textsc{latent}|X{_i}(n-1) = \textsc{susceptible})$, the probability that a susceptible individual is infected at $n$. Each individual belongs to a number of mixing groups with which an agent interacts, denoted $g \in \mathcal{G}_i(n)$, as well as an associated static set of agents $\mathcal{A}_g$. We define {the context-dependent probability} $p^g_{j \rightarrow i}$ that {infectious} individual $j$ infects {susceptible} individual $i$ in context $g$ in a single time step. The probability that a susceptible agent $i$ is infected at a given time step $n$ is thus calculated as:
\begin{equation}
    p_i(n) = 1 - \prod_{g \in G_i(n)} \left[  \prod_{j \in \mathcal{A}_g\setminus i} (1 - p^g_{j \rightarrow i}(n))  \right].
		\label{eq}
\end{equation}
We also define a scaling factor $\kappa$ (proportional to the reproductive number $R_0$), as a free parameter which allows us to vary the contagiousness of simulated epidemic scenarios:
\begin{equation} \label{eq:prob-transmission}
	p_{j \to i}^g(n ) = \kappa \ f( n - n_j \mid j ) \ q_{j \to i}^g
\end{equation}
where $n_j$ denotes the time when agent $j$ becomes infected, and $q_{j \to i}^g$ is the probability of transmission from agent $j$ to $i$ at the infectivity peak, derived from the  transmission or contact rates. { The function $f : \mathbb{N} \to [0,1]$ represents the infectivity of case $j$ over time: $f( n - n_j \mid j ) = 0$ when $n < n_j$, implementing the profile shown in Supplementary Fig.~\ref{fig:nhd}.}

This model assumes that for all contexts, the probabilities of infection over a given time period are known. In cases where this information is unavailable, we instead utilise contact rates reported and calibrated in previous studies. Thus, a majority of the transmission and contact probabilities follow  previous work on pandemic influenza~\cite{halloran2008modeling,mossong2008social,chao2010flute,cauchemez2011role,Cliff2018,Zachreson2018}, see Supplementary Tables~\ref{tab:contact-table} and~\ref{tab:transmission-rates} in Appendix~\ref{rates}. Full details regarding their application can be found in~\cite{Cliff2018}.

In this study we used ``the attack rate pattern weighted index case'' method to calculate $R_0$~\cite{GermannKadauEtAl2006,Zachreson2020}. The method is based on age-specific attack rates, computed as averages over many simulation instances, in order to reduce the bias in determining a typical index case, present due to population heterogeneity. As argued in~\cite{miller2009spread,Zachreson2020}, given the correlation between age group and population structure, the age-stratified weights, assigned to secondary cases produced by a sample of index cases, improve the estimation of the reproductive number $R_0$. Possible outliers were removed by the interquartile (IQR) method, using 1.5 IQR rule, which makes no assumptions about an underlying distribution. Five age groups were used: [0--4, 5--18, 19--29, 30--64, 65+], with the age-dependent attack rates [0.02, 0.04, 0.18, 0.58, 0.18], producing the corresponding age-dependent reproductive numbers: [1.16, 3.44, 2.63, 2.65, 3.35], with the weighted average of the adjusted reproductive number $R_0 = 2.77$, with 95\% CI [2.73, 2.83], constructed from the bias corrected bootstrap distribution (sample size 6,315).

\section{Results of sensitivity analysis}
\label{sens-res}

\subsection{Sensitivity of the model}
\label{sens-res1}

\bgroup
\begin{table}
	\caption{The input parameters $x_i$ and output variables $y_j$: local sensitivity analysis with the response $|F_{i,j}|$. Source data are provided as Supplementary Data 1.}
	\label{tab:sensit}
	\vspace{1mm}
	\centering
\resizebox{1.0\textwidth}{!}{%
	{\raggedright
	 \noindent
	\setlength\tabcolsep{4pt} 
	{ 
	 \begin{tabular}{lllcccccccc}
	Parameter & Default & Range & \multicolumn{2}{c}{$R_0$ ($y_1$)} & \multicolumn{2}{c}{$T_{gen}$ ($y_2$)} & \multicolumn{2}{c}{$\dot{C}$ ($y_3$)} & \multicolumn{2}{c}{$A_c$ ($y_4$)} \\
	& & & $\mu^*_{i,1}$ & $\sigma_{i,1}$ & $\mu^*_{i,2}$ & $\sigma_{i,2}$ & $\mu^*_{i,3}$ & $\sigma_{i,3}$ & $\mu^*_{i,4}$ & $\sigma_{i,4}$  \\
	\hline
 Time-to-peak, days ($x_1$) & 5 & $[4, 7]$ & 0.47 & 0.61 & 1.32 & 1.27 & 0.09 & 0.01 & 0.002 & 0.0002 \\	
 Recovery period, days ($x_2$) & 12 & $[7, 21]$ & 2.79 & 1.13 & 5.50 & 0.66 & 0.08 & 0.01 & 0.014 & 0.0005 \\
 Asymptomatic infectivity ($x_3$) & 0.3 & $[0.05, 0.45]$ & 0.69 & 0.75 & 2.42 & 1.48 & 0.10 & 0.01 & 0.004 & 0.0004 \\
 Symptomatic adults ($x_4$) & 0.669 & $[0.5, 0.8]$ & 0.83 & 0.60 & 0.65 & 0.54 & 0.06 & 0.01 & 0.033 & 0.0001 \\
 Symptomatic children ($x_5$) & 0.134 & $[0.05, 0.25]$ & 0.43 & 0.55 & 0.38 & 0.45 & 0.07 & 0.01 & 0.085 & 0.0001 \\
\hline
	\end{tabular}
	} 
	}
}
\end{table}
\egroup

\bgroup
\begin{table}
	\caption{The input parameters $x_i$ and output variables $y_j$: global sensitivity analysis with the effect $|F_{i,j}|$. Source data are provided as Supplementary Data 2.}
	\label{tab:sensitGSA}
	\vspace{1mm}
	\centering
\resizebox{0.95\textwidth}{!}{%
	{\raggedright
	 \noindent
	\setlength\tabcolsep{4pt} 
	 \begin{tabular}{llcccccccc}
	Parameter &  Range & \multicolumn{2}{c}{$R_0$ ($y_1$)} & \multicolumn{2}{c}{$T_{gen}$ ($y_2$)} & \multicolumn{2}{c}{$\dot{C}$ ($y_3$)} & \multicolumn{2}{c}{$A_c$ ($y_4$)} \\
	& & $\mu^*_{i,1}$ & $\sigma_{i,1}$ & $\mu^*_{i,2}$ & $\sigma_{i,2}$ & $\mu^*_{i,3}$ & $\sigma_{i,3}$ & $\mu^*_{i,4}$ & $\sigma_{i,4}$  \\
	\hline
 Time-to-peak, days ($x_1$) &  $[4, 7]$ & 0.22 & 0.14 & 0.98 & 0.26 & 0.04 & 0.01 & 0.003 & 0.0038 \\	
 Recovery period, days ($x_2$) &  $[7, 21]$ & 2.73 & 0.30 & 5.12 & 0.48 & 0.05 & 0.02 & 0.015 & 0.0165 \\
 Asymptomatic infectivity ($x_3$)  & $[0.05, 0.45]$ & 0.88 & 0.41 & 2.19 & 0.81 & 0.08 & 0.02 & 0.005 & 0.005 \\
 Symptomatic adults ($x_4$) &  $[0.5, 0.8]$ & 0.94 & 0.36 & 0.89 & 0.55 & 0.02 & 0.01 & 0.042 & 0.027 \\
 Symptomatic children ($x_5$) &  $[0.05, 0.25]$ & 0.10 & 0.11 & 0.12 & 0.09 & 0.02 & 0.01 & 0.089 & 0.0117 \\
\hline
	\end{tabular}
	}
}
\end{table}
\egroup

Results of the local sensitivity analysis are summarised in Supplementary Table~\ref{tab:sensit}. The analysis shows that the mean values $\mu^*_{2,1}$ and $\mu^*_{2,2}$, measuring the influence of the recovery period on $R_0$ and $T_{gen}$ respectively, are larger than the means of the other responses $|F_{i,1}|$ and $|F_{i,2}|$.  This indicates that the reproductive ratio and the generation period are most sensitive to changes in the recovery period ($x_2$). 
Given the range of the recovery period, varied between 7 and 21 days, each discretisation step $\Delta = 0.1$ corresponds to the recovery period's change of 1.4 days. For each 1.4-day variation, the reproductive ratio changes by 0.279 on average, resulting in the mean response value $\mu^*_{2,1} = 2.79$. Over the ten steps, these variations extend the reproductive ratio by approximately 0.2 to 0.4 per step, from $R_0 = 1.81$ ($x_2 = 7$ days) to $R_0 = 4.59$ ($x_2 = 21$ days), mostly linearly, as shown in Supplementary Fig.~\ref{FigS}.b.
Similarly, these variations (linearly) extend the generation period by approximately 0.4 to 0.6 per step, from $T_{gen} = 5.51$ ($x_2 = 7$ days) to $T_{gen} = 11.01$ ($x_2 = 21$ days), Supplementary Fig.~\ref{FigS}.b.  For each discretisation step, the estimates of $R_0$ and $T_{gen}$ are produced by ``the attack rate pattern weighted index case'' method described in section~\ref{trans}, with $n = 6,655$ runs on average.

Changes in the other input parameters result in smaller effects on $R_0$ and $T_{gen}$, as shown in Supplementary Fig.~\ref{FigS}. 
Overall, despite the sensitivity of the first two output variables to changes in the recovery period ($x_2$), their variations are within the expected ranges, demonstrating robustness of the model in terms of the reproductive ratio and the generation period. 

The other two output variables, the daily growth rate of cumulative incidence at day 50, $\dot{C}$, and the attack rate in children, $A_c$, show small sensitivity to all input parameters, indicated by the low means $\mu^*_{i,3}$ and $\mu^*_{i,4}$. The asymptomatic infectivity ($x_3$) is the parameter influencing the growth rate $\dot{C}$ slightly more than other inputs. In response to varying $x_3$, the daily growth rate at day 50 changes between 0.11 and 0.19, which is an acceptable range, see Supplementary Fig.~\ref{FigS}.c. Not surprisingly, the fraction of symptomatic cases among children ($x_5$) is the parameter with the highest effect on the children attack rate $A_c$.  When $x_5$ is varied, the attack rate in children changes between 2\% and 11\%, again within an acceptable range, see Supplementary Fig.~\ref{FigS}.e.  Thus, all input parameters are weakly influential with respect to output variables $y_3$ and $y_4$.

Supplementary Figure~\ref{FigS} shows results of the sensitivity analysis of the model, in terms of five input parameters: the time-to-peak (days, $x_1$), the recovery period (days, $x_2$), the probability of transmission for asymptomatic agents. i.e., asymptomatic infectivity ($x_3$), the fraction of symptomatic cases in adults, i.e., symptomatic adults ($x_4$), and the fraction of symptomatic cases in children, i.e., symptomatic children ($x_5$). For each of the input parameters, we trace two output variables which are most affected by this specific input, selected based on ranking of responses $|F_{i,j}|$ using the means $\mu^*_{i,j}$, as reported in Supplementary Table~\ref{tab:sensit}.  These dependencies are mostly linear, and the output variables are bounded within their anticipated ranges, indicating robustness of the model. 

\begin{figure}[ht]
    \centering
    \includegraphics[clip, trim=7.9cm 2.2cm 7.9cm 0cm, width = 1.0\linewidth]{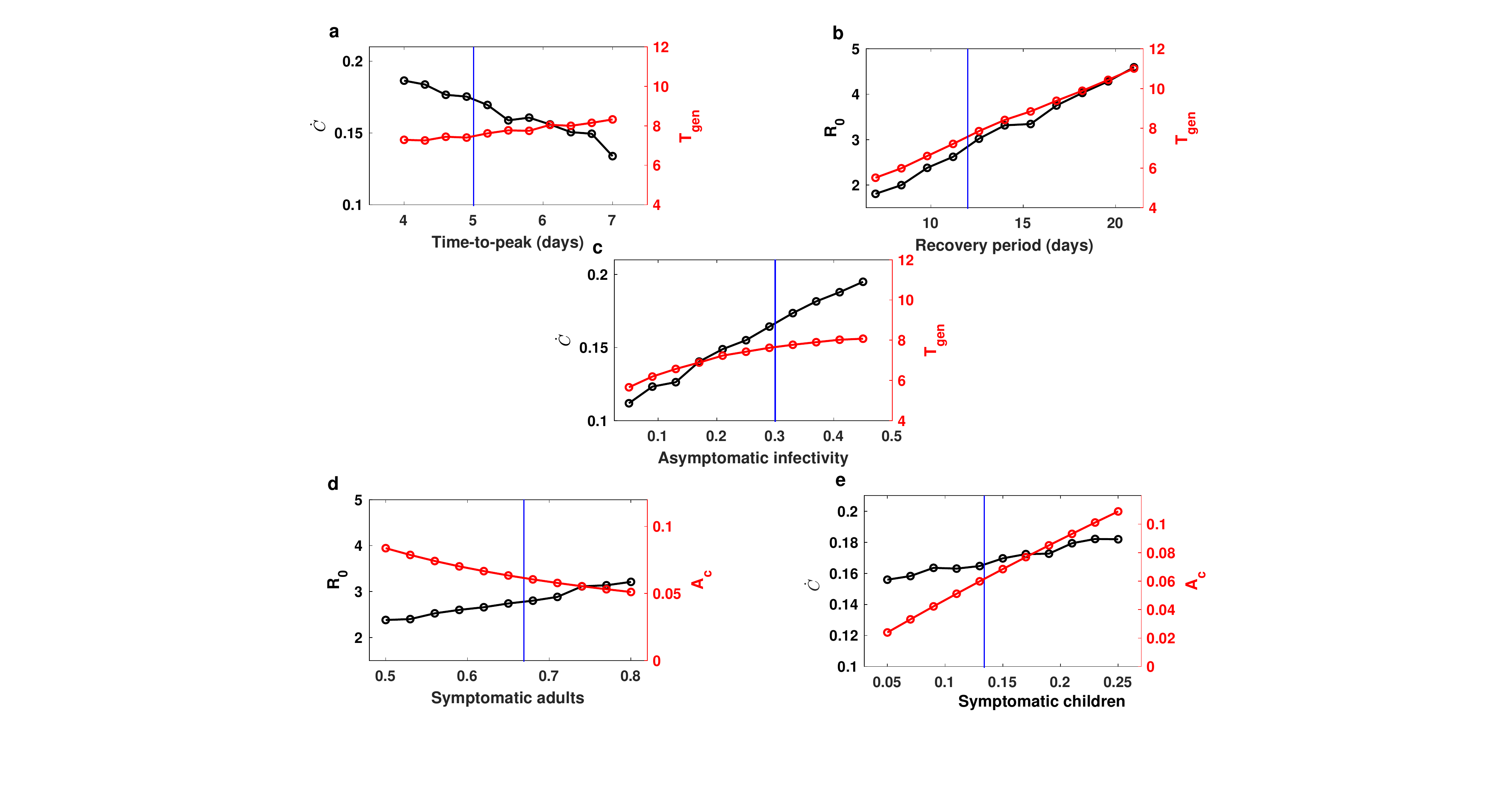}
    \caption{\textbf{Local sensitivity dependencies of the two most affected output variables for each input parameter.}  For each $x_i$, the output variables $y_j$ with the two highest mean values $\mu^*_{i,j}$ are shown, cf. Supplementary Table~\ref{tab:sensit}. The default value of each input parameter is shown with a vertical line. Source data are provided as Supplementary Data 1.}
    \label{FigS}
\end{figure}

Supplementary Table~\ref{tab:sensitGSA} summarises results of the global sensitivity analysis using the Morris method, carried out with $r = 20$ repeats and $k = 5$ inputs, resulting in 120 parameter combinations, i.e., $r (k + 1)$, with inputs varied over $l = 10$ discretisation levels. When estimating $R_0$ and $T_{gen}$ for each parameter combination, we use ``the attack rate pattern weighted index case'' method described in section~\ref{trans}, with $n = 6,702$ runs on average. For other two output variables $\dot{C}$ and $A_c$, we run simulations $m = 10$ times for each parameter combination,  averaging the results over these runs before computing the sensitivity effects. 

In concordance with the LSA, the reproductive ratio $R_0$ and the generation period $T_{gen}$ are most sensitive to changes in the recovery period, but also stay within the expected ranges (e.g., $R_0$ varies between 1.33 and 4.96, and $T_{gen}$ varies between 3.78 and 11.79).  The daily growth rate of cumulative incidence at day 50, $\dot{C}$, and the attack rate in children, $A_c$, show small global sensitivity to all input parameters, despite strong parameter interactions affecting $A_c$, as  evidenced by the higher global $\sigma_{i,4}$.

In summary, the analysis shows that the model is robust to changes in the input parameters, with the highest sensitivity detected in the reproductive ratio and the generation period, in response to the recovery period. Even for the most affected variables, the resulting variations are limited within their expected ranges.

\subsection{Sensitivity of the model outcomes}
\label{sens-res2}

We also investigate whether the model outcomes are sensitive with respect to three  
context-dependent micro-distancing levels: within households, community, and workplace/school environments. 
Two specific targets are considered:
\begin{enumerate}[label=(\roman*),noitemsep]
\item the epidemic dynamics traced along $90\%$ SD compliance,
\item the transition across the levels of SD compliance, in the range between 70\% and 80\% levels.
\end{enumerate}
The epidemic dynamics traced along $90\%$ SD compliance is of primary relevance (cf. Fig.~3 of the main manuscript), and so sensitivity of the corresponding outcomes, registered at the end of suppression, is important to establishing the applicability range of the model. The  transition in the range between 70\% and 80\% levels is our main policy-informing result (cf. Fig.~4 of the main manuscript), and its robustness is crucial for our study.
For each target, the output variables of interest include the prevalence and cumulative incidence, registered at the end of simulated suppression period. 

Each level of micro-distancing is varied by 5\% within a 50\% range, in proximity of the default values, specified in Supplementary Tables~\ref{tab:sensit2} and~\ref{tab:sensit3}. Using discretisation step $\Delta = 0.1$ and 10 runs per step, we compute the corresponding responses of prevalence and cumulative incidence under 90\% SD compliance (Supplementary Table~\ref{tab:sensit2}), as well as the responses of the difference  $\ominus$ between the outcomes under 70\% and 80\% SD compliance, for both prevalence and cumulative incidence (Supplementary Table~\ref{tab:sensit3}).

\bgroup
\begin{table}
	\caption{Local sensitivity of the epidemic dynamics under $90\%$ SD compliance to micro-distancing levels. Source data are provided as Supplementary Data 1.}
	\label{tab:sensit2}
	\vspace{1mm}
	\centering
\resizebox{0.9\textwidth}{!}{%
	{\raggedright
	 \noindent
	\setlength\tabcolsep{4pt} 
	{ 
	 \begin{tabular}{lllcccc}
	Level of & Default & Range & \multicolumn{2}{c}{Prevalence} & \multicolumn{2}{c}{Cumulative Incidence} \\
	micro-distancing in & & & $\mu^*, \times 10^3$ & $\sigma, \times 10^3$ & $\mu^*, \times 10^3$ & $\sigma, \times 10^3$   \\
	\hline 
 households & 100\% & $[75\%, 125\%]$ & 0.400 & 0.061 & 7.919 & 1.009 \\	
 community & 50\% & $[25\%, 75\%]$ & 2.449 & 0.291 & 15.966 & 1.895 \\
 workplace/schools & 0\% & $[0\%, 50\%]$ & 3.168 & 2.157 & 84.106 & 5.017 \\
\hline
	\end{tabular}
	} 
	}
}
\end{table}
\egroup

\bgroup
\begin{table}
	\caption{Local sensitivity of the transition between 70\% and 80\% SD levels to micro-distancing levels. Source data are provided as Supplementary Data 1.}
	\label{tab:sensit3}
	\vspace{1mm}
	\centering
\resizebox{0.9\textwidth}{!}{%
	{\raggedright
	 \noindent
	\setlength\tabcolsep{4pt} 
	{ 
	 \begin{tabular}{lllcccc}
	Level of & Default & Range & \multicolumn{2}{c}{$\ominus$ Prevalence} & \multicolumn{2}{c}{$\ominus$ Cumulative Incidence} \\
	micro-distancing in & & & $\mu^*, \times 10^3$ & $\sigma, \times 10^3$ & $\mu^*, \times 10^3$ & $\sigma, \times 10^3$   \\
	\hline 
 households & 100\% & $[75\%, 125\%]$ & 11.815 & 1.075 & 38.739 & 3.901 \\	
 community & 50\% & $[25\%, 75\%]$ & 16.089 & 2.511 & 46.822 & 5.539 \\
 workplace/schools & 0\% & $[0\%, 50\%]$ & 175.140 & 26.454 & 327.180 & 51.540 \\
\hline
	\end{tabular}
	} 
	}
}
\end{table}
\egroup

\bgroup
\begin{table}
	\caption{Global sensitivity of the epidemic dynamics under $90\%$ SD compliance to micro-distancing levels. Source data are provided as Supplementary Data 2.}
	\label{tab:sensit2GSA}
	\vspace{1mm}
	\centering
\resizebox{0.85\textwidth}{!}{%
	{\raggedright
	 \noindent
	\setlength\tabcolsep{4pt} 
	 \begin{tabular}{llcccc}
	Level of &  Range & \multicolumn{2}{c}{Prevalence} & \multicolumn{2}{c}{Cumulative Incidence} \\
	micro-distancing in &  & $\mu^*, \times 10^3$ & $\sigma, \times 10^3$ & $\mu^*, \times 10^3$ & $\sigma, \times 10^3$   \\
	\hline 
 households &  $[75\%, 125\%]$ & 0.463 & 0.670 & 4.550 & 3.020 \\	
 community &  $[25\%, 60\%]$ & 0.951 & 0.658 & 8.026 & 3.397 \\
 workplace/schools &  $[0\%, 25\%]$ & 0.900 & 1.067 & 5.980 & 3.937 \\
\hline
	\end{tabular}
	}
}
\end{table}
\egroup

\bgroup
\begin{table}
	\caption{Global sensitivity of the transition between 70\% and 80\% SD levels to micro-distancing levels. Source data are provided as Supplementary Data 2.}
	\label{tab:sensit3GSA}
	\vspace{1mm}
	\centering
\resizebox{0.85\textwidth}{!}{%
	{\raggedright
	 \noindent
	\setlength\tabcolsep{4pt} 
	 \begin{tabular}{llcccc}
	Level of &  Range & \multicolumn{2}{c}{$\ominus$ Prevalence} & \multicolumn{2}{c}{$\ominus$ Cumulative Incidence} \\
	micro-distancing in &  & $\mu^*, \times 10^3$ & $\sigma, \times 10^3$ & $\mu^*, \times 10^3$ & $\sigma, \times 10^3$   \\
	\hline 
 households &  $[75\%, 125\%]$ & 6.162 & 4.740 & 14.732 & 12.136 \\	
 community &  $[25\%, 60\%]$ & 10.982 & 4.530 & 20.825 & 8.769 \\
 workplace/schools &  $[0\%, 25\%]$ & 17.749 & 8.248 & 38.416 & 16.504 \\
\hline
	\end{tabular}
	}
}
\end{table}
\egroup

We conclude that both targets are much more sensitive to variations in the micro-distancing levels within the workplace/school environments, and least sensitive to micro-distancing within households. Importantly, the sensitivity dependencies are linear around the default values of input parameters, as shown in Supplementary Figures~\ref{FigSD} and~\ref{FigSD2}. This indicates that the model outcomes quantifying the contribution of social macro-distancing to the disease control are robust to the levels of micro-distancing, within certain levels. The onset of non-linearity, seen in Supplementary Figures~\ref{FigSD}.b and~Fig.~\ref{FigSD}.c, as well as in Supplementary Figures~\ref{FigSD2}.b and~\ref{FigSD2}.c, marks the range of applicability in terms of the corresponding micro-distancing parameters. Specifically, the levels of micro-distancing within the workplace/school environments should not exceed 25\% (Supplementary Figures~\ref{FigSD}.c and~\ref{FigSD2}.c), and within the community should stay below 60\% (Supplementary Figures~\ref{FigSD}.b and~\ref{FigSD2}.b), as going beyond these levels increases the sensitivity of the results in a non-linear fashion.

This is further confirmed by the global sensitivity analysis carried out using the Morris method, applied to reduced parameter ranges for the community and workplace/school environments. For each target, the analysis uses $r = 20$ repeats and $k = 3$ inputs (varied over $l = 10$ discretisation levels), resulting in $80 = r (k + 1)$ parameter combinations, each simulated $m = 10$ times. These results, summarised in Supplementary Tables~\ref{tab:sensit2GSA} and~\ref{tab:sensit3GSA}, show that the targets are least sensitive to micro-distancing within households. Another notable observation is that there are limited interactions among micro-distancing parameters, as evidenced by moderate values of $\sigma$. Importantly, the model outcomes are robust to globally varying micro-distancing in all social contexts, when these variations are within the identified ranges of applicability.

\begin{figure}[ht]
    \centering
    \includegraphics[clip, trim=7.6cm 2.2cm 8.0cm 1cm, width = 1.0\linewidth]{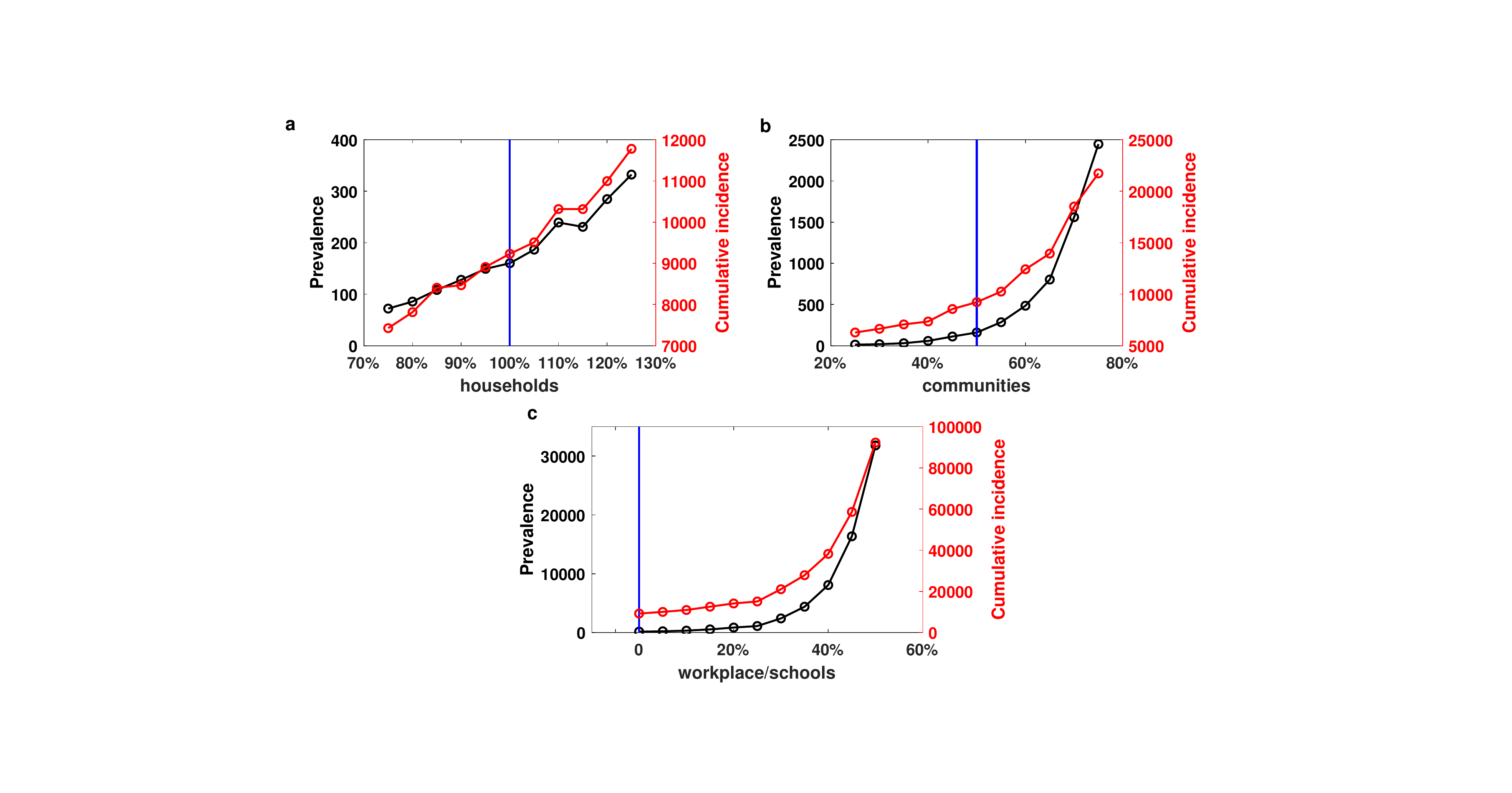}
    \caption{\textbf{Local sensitivity of prevalence and cumulative incidence.} The sensitivity is traced for $90\%$ SD compliance, coupled with case isolation, home quarantine, and international travel restrictions, to changes in  micro-distancing levels in \textbf{a}~households, \textbf{b} community, \textbf{c} workplace/school environments. The default value of each input parameter is shown with a vertical line. Source data are provided as Supplementary Data 1.}
    \label{FigSD}
\end{figure}

\begin{figure}[ht]
    \centering
    \includegraphics[clip, trim=7.6cm 2.2cm 8.0cm 1cm, width = 1.0\linewidth]{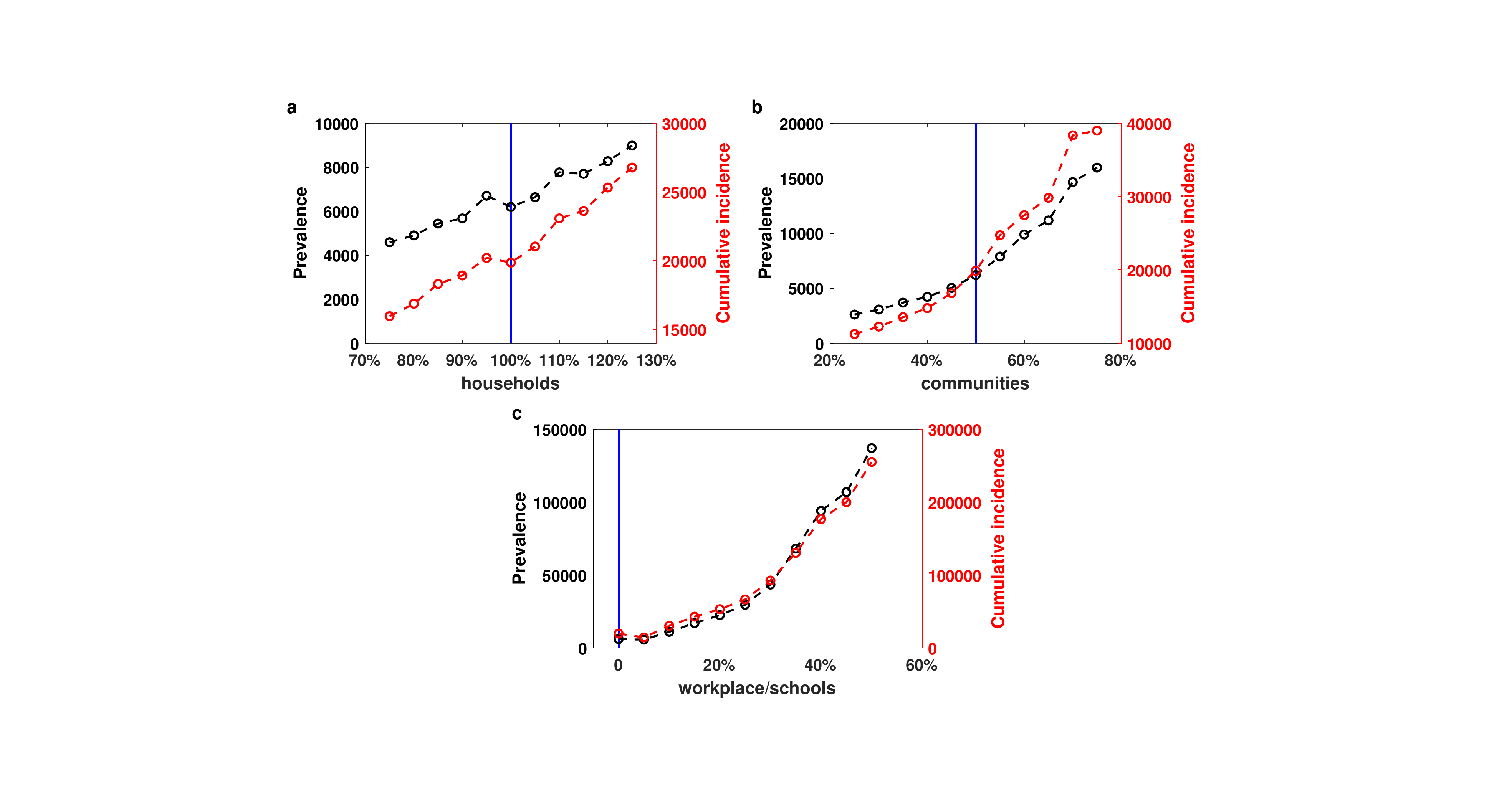}
    \caption{\textbf{Local sensitivity of changes $\ominus$ in prevalence and cumulative incidence}. The sensitivity is traced across 70\% and 80\% SD levels, coupled with case isolation, home quarantine, and international travel restrictions, to changes in  micro-distancing levels in \textbf{a} households, \textbf{b} community, \textbf{c}~workplace/school environments. The default value of each input parameter is shown with a vertical line. Source data are provided as Supplementary Data 1.}
    \label{FigSD2}
\end{figure}

\section{Transmission and contact probabilities}
\label{rates}
Following \cite{Cliff2018}, with some minor adjustments, the { contact and transmission} probabilities are given in Supplementary Tables~\ref{tab:contact-table} and~\ref{tab:transmission-rates}, respectively.

\bgroup
\begin{table}
	\caption{Daily contact probabilities $c_{j \to i}^g$ for different contact groups $g$, reported by~\cite{chao2010flute}, reproduced from \cite{Cliff2018}, except for the rates in household clusters. The age is assigned an integer value.}
	\label{tab:contact-table}
	\vspace{1mm}
	\centering
\resizebox{1.0\textwidth}{!}{%
	{\raggedright
	 \noindent
	\begin{tabular}{llll}
	Mixing group $g$ & Infected individual $j$ & Susceptible individual $i$ & Contact probability $c_{j \to i}^g$ \\
	\hline
	Household cluster & Child ($\le 18$) & Child ($\le 18$) &  0.05 \\
	 & Child ($\le 18$) & Adult ($\ge 19$) & 0.05 \\
	 & Adult ($\ge 19$) & Child ($\le 18$) & 0.05  \\
	 & Adult ($\ge 19$) & Adult ($\ge 19$) & 0.05 \\
	\hline
	Working Group & Adult (19-64) & Adult (19-64) & 0.05 \\
	\hline
	Neighbourhood & Any & Child (0-4) &  0.0000435 \\
	 & Any & Child (5-18) & 0.0001305 \\
	 & Any & Adult (19-64) & 0.000348  \\
	 & Any & Adult ($\ge 65$) & 0.000696 \\
	\hline
	Community & Any & Child (0-4) &  0.0000109 \\
	 & Any & Child (5-18) & 0.0000326 \\
	 & Any & Adult (19-64) & 0.000087  \\
	 & Any & Adult ($\ge 65$) & 0.000174 \\
	\hline
	\end{tabular}
	}
}
\end{table}
\egroup

\bgroup
\def\arraystretch{1.3}
\setlength\arrayrulewidth{1pt}
\setlength\tabcolsep{4mm}
\begin{table}
\centering
 \noindent
 \caption{Daily transmission probabilities $q_{j\to i}^g$ for different contact groups $g$, reported by~\cite{cauchemez2011role}, reproduced from \cite{Cliff2018}. The age is assigned an integer value.}
 \label{tab:transmission-rates}
 \vspace{1mm}
 \resizebox{1.0\textwidth}{!}{%
 	{\raggedright
\begin{tabular}{llll}
Contact Group $g$ & Infected Individual $j$ & Susceptible Individual $i$ & Transmission Probability $q^g_{j \to i}$ \\
\hline

Household size 2 & Any & Child ($\le 18$) & 0.0933 \\
& Any & Adult ($\ge 19$) & 0.0393 \\
\hline
Household size 3 & Any &  Child ($\le 18$) & 0.0586 \\
 & Any & Adult ($\ge 19$) & 0.0244 \\
\hline
Household size 4 & Any & Child ($\le 18$) & 0.0417 \\
& Any & Adult ($\ge 19$) & 0.0173 \\
\hline
Household size 5 & Any & Child ($\le 18$) & 0.0321 \\
 & Any & Adult ($\ge 19$) & 0.0133 \\
\hline
Household size 6 & Any & Child ($\le 18$) & 0.0259 \\
& Any &  Adult ($\ge 19$) & 0.0107 \\
\hline
School & Child ($\le 18$) & Child ($\le 18$) & 0.000292 \\
Grade & Child ($\le 18$) & Child ($\le 18$) & 0.00158 \\
Class & Child ($\le 18$) & Child ($\le 18$) & 0.035 \\
\hline
\end{tabular}
}}
\label{transmission_table}
\end{table}
\egroup

\vspace*{-10mm}
\section{Population generation, demographics and mobility}
\label{demogr}
Prior to the { \amtrac} simulations, a surrogate population is generated to match coarse-grained distributions arising from the 2016 Australian census, published by the Australian Bureau of Statistics (ABS). In generating this surrogate population, we use Statistical Areas (SA1 and SA2) level statistics, comprising age, household composition and workplaces. Individuals in the population are separated into 5 different age groups: preschool aged children (0-4), children (5-18), young adults (19-29), adults (30-65) and older adults (65+). Along with these assigned characteristics, individuals are assigned a number of mixing contexts based on the census data. The model uses a discrete-time simulation, where each simulated day is separated into two distinct portions: `daytime' and `nighttime'. In the daytime, workplace and school-based mixing are considered, whereas nighttime mixing considers the household transmissions, as well as other local spread at the neighborhood (SA1) and community (SA2) levels. 

The population generation begins with the contexts needed for nighttime mixing, which can be thought of as ``home regions''. The simulation iterates through each SA1, creating a cumulative density function (CDF) describing the size and type of households, based on two dependent probability distributions defined by the ABS. Given this CDF, the procedure begins to randomly generate households, with the generation of agents occurring during this process. Once a household is generated for an SA1, agents are generated to match the size and type of the household (e.g., a single parent family of size four will generate one adult and three children). In order to generate attributes for this surrogate population, the simulation then reads in CDFs describing the population statistics of the given SA, with each of these agents being assigned some attributes based on these population distributions.

Following the population of the home regions, the simulator assigns work and school regions to individuals within the population. This process is based on the ``Travel to work'' data published by the ABS, which defines a number of individuals $N$  living in home region $i$ and working in region $j$. In order to satisfy each of these ``worker flows'', a number of unassigned working-age individuals (19-64 years old) in region $i$ is selected at random and assigned to work in location $j$. School allocation, on the other hand, is somewhat more complicated as the detailed data about student home locations are not available from the ABS. Instead, we use the available data from the Australian Curriculum, Assessment and Reporting Authority (ACARA), detailing the locations of schools, along with a proximity based model which biases children allocation towards closer schools. More detail about student allocation can be found in previous studies~\cite{Zachreson2018}.

\vspace*{-2mm}
\section{Effects of school closures}	
\label{schools}

\begin{figure}
	\centering
  \includegraphics[clip, trim=8.0cm 0cm 7.5cm 0cm, width=1.0\textwidth]{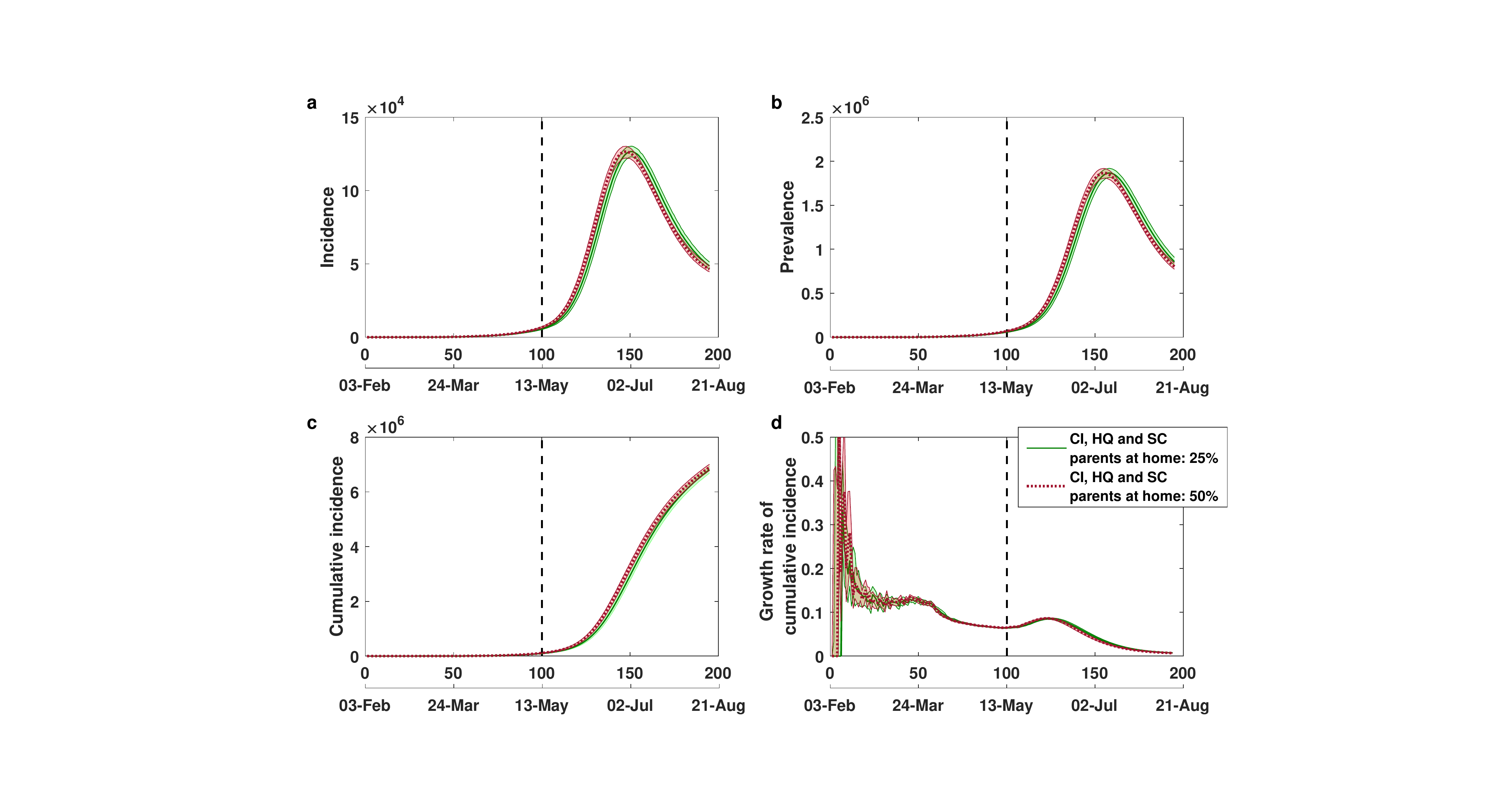}
\caption{\textbf{Effects of parents' commitment to stay home during school closures.} Increasing parents' commitment to stay home during school closures (SC) from 25\% (solid) to 50\% (dashed) does not significantly affect the spread: \textbf{a} incidence, \textbf{b}~prevalence, \textbf{c} cumulative incidence, \textbf{d} daily growth rate of cumulative incidence, shown as average (solid) and 95\% confidence interval (shaded) profiles, over 20 runs. The 95\% confidence intervals are constructed from the bias corrected bootstrap distributions. The strategy with school closures (SC) combined with case isolation {(CI)} and home quarantine {(HQ)} lasts 49 days (7 weeks), marked by a vertical dashed line. 
Restrictions on international arrivals are set to last until the end of each scenario.  The alignment between simulated days and actual dates may slightly differ across separate runs.}
\label{SC_comparison}
\end{figure}

\begin{figure}
	\centering
  \includegraphics[clip, trim=7.7cm 0cm 8.0cm 0cm, width=1.0\textwidth]{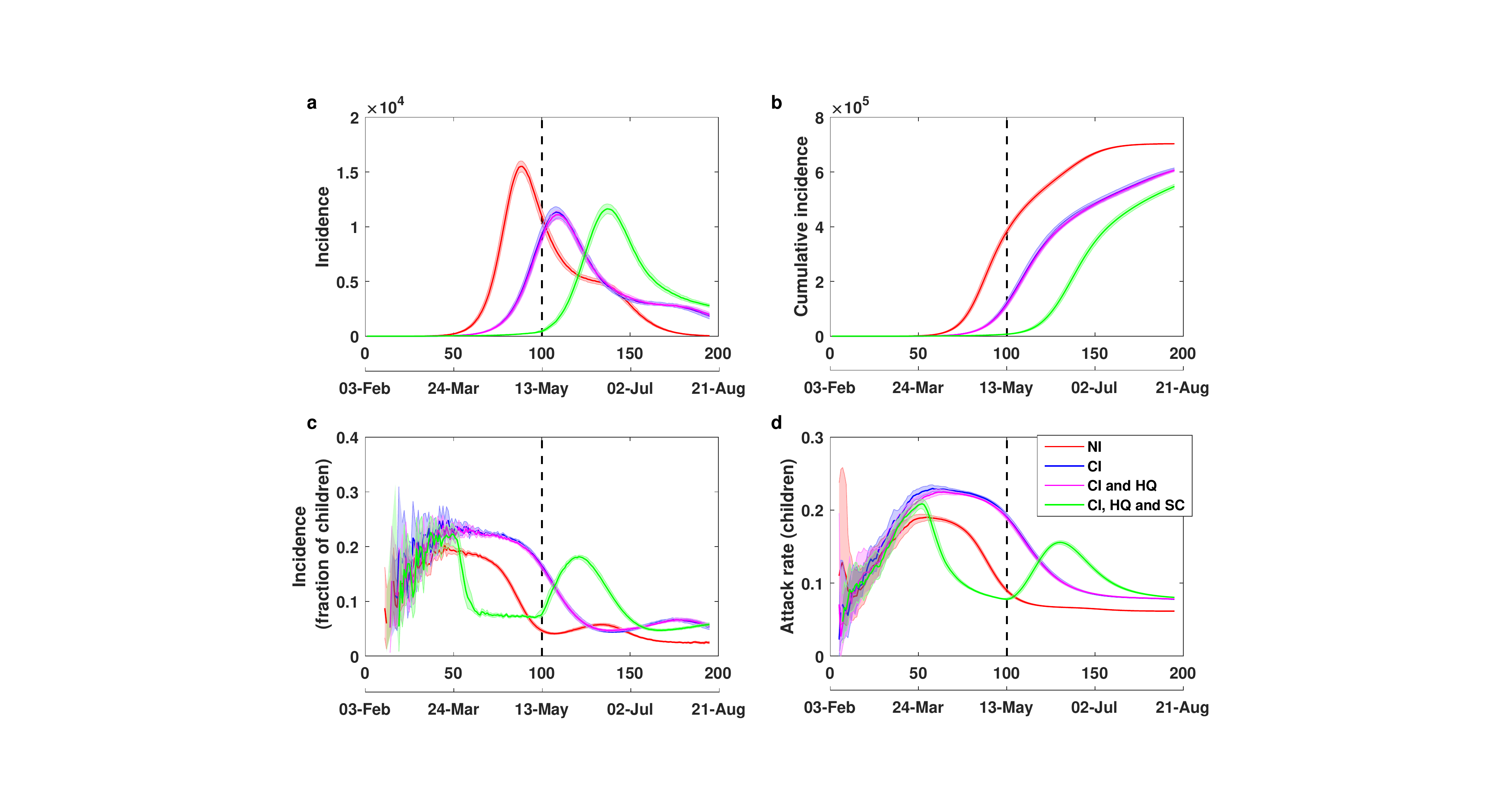}
\caption{\textbf{Effects of school closures: children.} School closures {(SC)} delay incidence peak by four weeks, but increase the fraction of new cases in children around the peak time by 7\%,  in comparison to case isolation (CI) and home quarantine (HQ), under international travel restrictions. No interventions: NI.  
Epidemic curves for children: \textbf{a} incidence, \textbf{b}~cumulative incidence, \textbf{c} fraction of children in incidence, and \textbf{d} fraction of children in cumulative incidence, shown as average (solid) and 95\% confidence interval (shaded) profiles, over 20 runs. The 95\% confidence intervals are constructed from the bias corrected bootstrap distributions. The strategy with school closures combined with case isolation and home quarantine lasts 49 days (7 weeks), marked by a vertical dashed line. 
Restrictions on international arrivals are set to last until the end of each scenario.  The alignment between simulated days and actual dates may slightly differ across separate runs.}
\label{SD_comparison_cld}
\end{figure}

\begin{figure}
	\centering
  \includegraphics[clip, trim=7.7cm 0cm 8.0cm 0cm, width=1.0\textwidth]{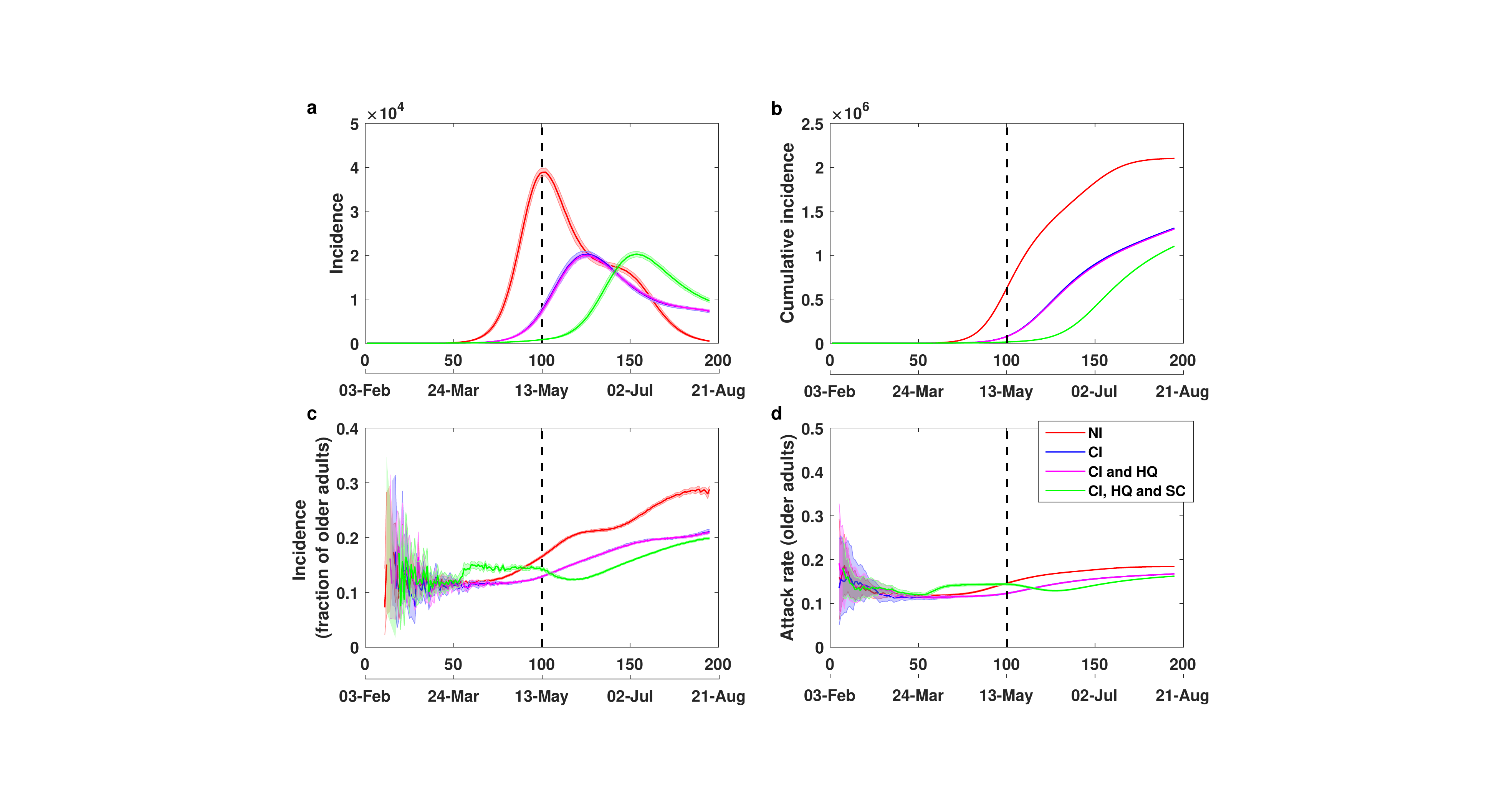}
\caption{\textbf{Effects of school closures: older adults.} School closures {(SC)} delay incidence peak by four weeks, but do not affect new cases for older adults,  in comparison to case isolation (CI) and home quarantine (HQ), under international travel restrictions. No interventions: NI. 
Epidemic curves for older adults: \textbf{a} incidence, \textbf{b} cumulative incidence, \textbf{c} fraction of older adults in incidence, and \textbf{d} fraction of older adults in cumulative incidence, shown as average (solid) and 95\% confidence interval (shaded) profiles, over 20 runs. The 95\% confidence intervals are constructed from the bias corrected bootstrap distributions. The strategy with school closures combined with case isolation and home quarantine lasts 49 days (7 weeks), marked by a vertical dashed line. 
Restrictions on international arrivals are set to last until the end of each scenario.  The alignment between simulated days and actual dates may slightly differ across separate runs.}
\label{SD_comparison_eld}
\end{figure}

Here we compare effects of school closures, added to the case isolation and home quarantine, for two levels of parents' commitment to stay home: 25\% and 50\%. That is, the proportion of children supervised at home during school closure by one of their randomly chosen parents varies from 25\% to 50\% (Supplementary Fig.~\ref{SC_comparison}).
Focussing on the 25\% commitment, we also trace the effects of school closures for two specific age groups: children and individuals over 65 years old (Supplementary Figures~\ref{SD_comparison_cld} and~\ref{SD_comparison_eld} respectively).

At this stage we revisit school closures in context of social distancing. As shown in Supplementary Fig.~\ref{School_compliance}, addition of the SC strategy to  SD set at 70\%  generates a reduction in incidence, albeit not lasting and progressing at a higher level than such reductions observed at  80\% and 90\% SD levels, coupled with school closures. This suggests that another potential but transient benefit of school closures is that it may ``compensate'' for about 10\% lack of SD compliance.

\begin{figure}
	\centering
  \includegraphics[clip, trim=8.6cm 1.5cm 9.0cm 1.5cm, width=1.0\textwidth]{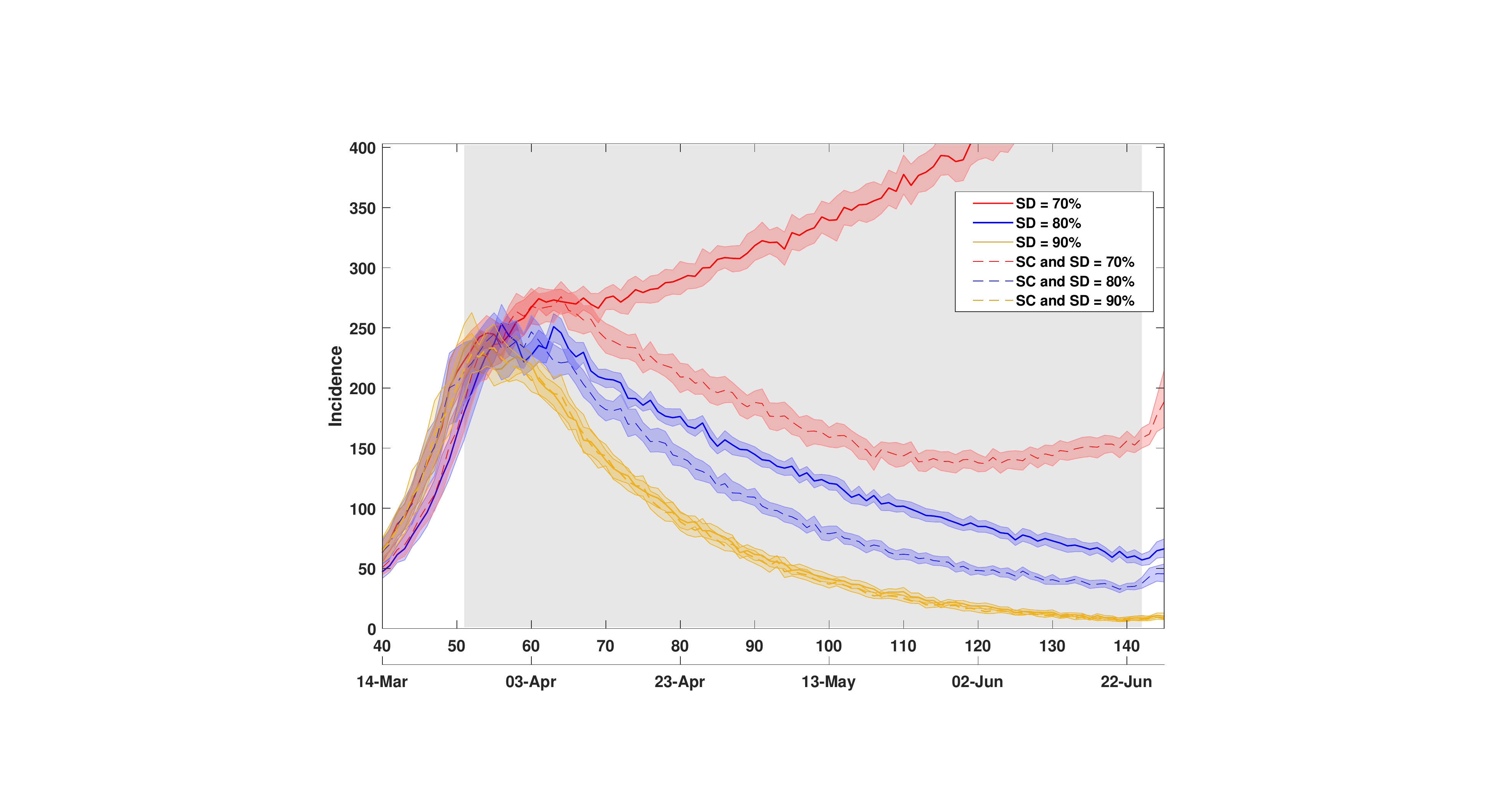}
\caption{\textbf{Effects of school closures combined with social distancing.}  School closures may temporarily ``compensate'' for about 10\% lack of social distancing (SD) compliance. A comparison of social distancing strategies, coupled with case isolation, home quarantine, international travel restrictions, and school closures (SC) or no school closures, across different compliance levels (70\%, 80\% and 90\%), shown as average (solid) and 95\% confidence interval (shaded) profiles, over 20 runs. The 95\% confidence intervals are constructed from the bias corrected bootstrap distributions. Duration of each combined SD and SC strategy is set to 91 days (13 weeks), shown as a grey shaded area. Case isolation, home quarantine  and restrictions on international arrivals are set to last until the end of each scenario.  The alignment between simulated days and actual dates may slightly differ across separate runs.}
\label{School_compliance}
\end{figure}

\vspace*{-2mm}
\section{Model validation}
\label{delay}

\subsection{A delayed introduction of strong social distancing measures}
\label{delay_comp}
On 21 March 2020, the number of confirmed COVID-19 cases in Australia crossed 1,000. This coincided with the ban on all international arrivals of non-residents, non-Australian citizens, put in place the night before. The primary scenario considered in this study introduces a social distancing policy, at varying degrees of compliance, triggered by crossing the threshold of 2,000 confirmed cases, exceeded in Australia three days later, on 24 March 2020, when strong measures (e.g., closures of non-essential services and places of social gathering) have been introduced. The primary scenario traced at $90\%$ SD, coupled with case isolation and home quarantine, is well-aligned with the actual epidemic timeline in Australia, as shown in Fig.~3 of the main manuscript, especially in terms of prevalence (Fig.~3.b) and cumulative incidence (Fig.~3.c). The actual daily incidence data (Fig.~3.a) are more noisy, having been affected, in particular, by separate clusters linked to infected cruise ships passengers. For example, by 18 April 2020, more than 600 COVID-19 cases in Australia, i.e., 10\% of total cases at the time, have been linked to the Ruby Princess cruise ship from which 2,700 passengers were allowed to disembark on 19 March~\cite{klein2020australia}. Despite the discrepancy in tracing the daily cases, our model accurately predicted timings of the incidence peak (Fig.~3.a) and prevalence peak (Fig.~3.b). In addition, the actual SD levels vary across time, and have been complemented by other surveillance, distancing and intervention measures, e.g., hotel quarantine of international arrivals, meticulous testing of health care workers, inter-state border closures, etc., which are not part of our model. To re-iterate, the model was calibrated by 24 March 2020, and the comparison across the SD levels pointed to $90\%$ SD as the closest match, but did not change the model parametrization, highlighting its robustness and predictive power.

\begin{figure}[ht]
\centering
    \includegraphics[clip, trim=8.0cm 0cm 6.6cm 0cm, width=1.0\textwidth]{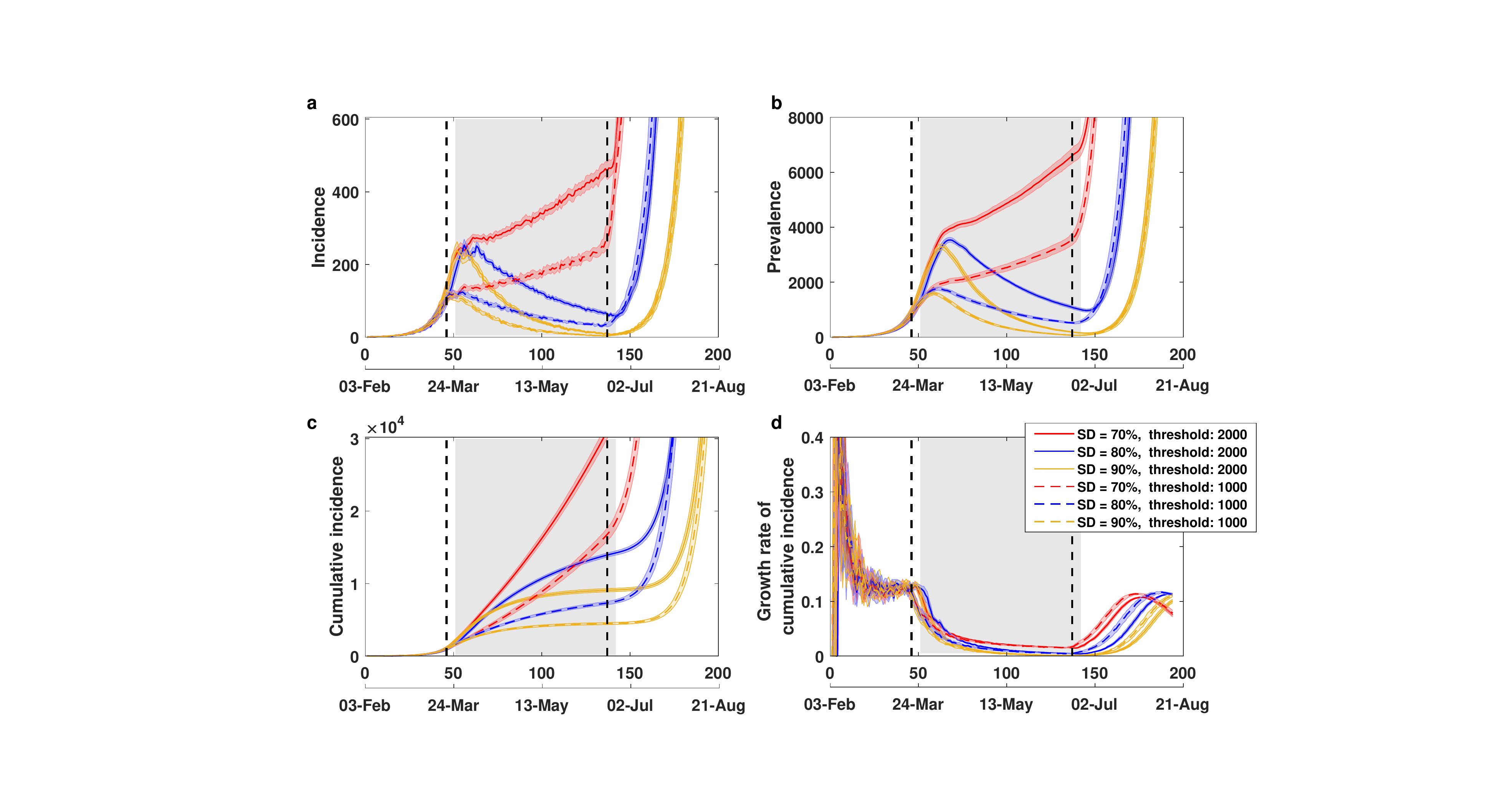}
\caption{\textbf{Effects of delays.} A three-day delay in introducing strict social distancing doubles the disease prevalence. A comparison of social distancing (SD) strategies, coupled with case isolation, home quarantine, {and international travel restrictions,} across different compliance levels (70\%, 80\% and 90\%). Two scenarios are contrasted: primary scenario with the threshold set at 2,000 cases (matching actual numbers on 24 March 2020), and the alternative threshold of 1,000 (matching actual numbers on 21 March 2020). Duration of each SD strategy is set to 91 days (13 weeks), shown as a grey shaded area for the primary threshold (2,000 cases), and with vertical dashed lines for the alternative threshold (1,000 cases). Case isolation, home quarantine, and restrictions on international arrivals are set to last until the end of each scenario. Traces include \textbf{a}~incidence, \textbf{b} prevalence, \textbf{c} cumulative incidence, and \textbf{d} the daily growth rate of cumulative incidence $\dot{C}$, shown as average (solid) and {95\% confidence interval (shaded) profiles, over 20 runs. The 95\% confidence intervals are constructed from the bias corrected bootstrap distributions. The alignment between simulated days and actual dates may slightly differ across separate runs.}}
    \label{SD_tri_com_preprint}
\end{figure}

To evaluate a delayed introduction of strong social distancing measures, we compare these two thresholds, separated by three days, while keeping all other parameters unchanged.
Referring to Supplementary Fig.~\ref{SD_tri_com_preprint}, a delayed response results in higher epidemic peaks, doubling the prevalence in comparison with the alternative scenario~{(Fig.~\ref{SD_tri_com_preprint}.b)}, across different levels of compliance. The cumulative incidence for $90\%$ SD nearly doubles as well, from around 5,000 total cases, to about 9,000~{(Supplementary Fig.~\ref{SD_tri_com_preprint}.c)}. 

We also observe that a three-day delay in introducing strong social distancing measures results in an approximately four-week lengthening of the required suppression period, confirmed by separate runs with a longer suppression duration (Supplementary Table~\ref{tab:delay}). The resultant difference (i.e., delay) averages in {23.56} days, with standard deviation of the difference estimated as {11.167} days.

\bgroup
\begin{table}
	\caption{The average differences between early and delayed interventions. A comparison between two scenarios: early interventions (threshold 1,000 cases) and delayed interventions (threshold 2,000 cases). Each scenario is evaluated over {20} runs.  In each run, a day is recorded when the prevalence decreases below a specified criterion in terms of active cases (ranging from 30 to 50 in increments of 5). Source data are provided as Source Data file.}
	\label{tab:delay}
	\vspace{1mm}
	\centering
\resizebox{.9\textwidth}{!}{%
	{\raggedright
	 \noindent
	{
	\begin{tabular}{ccccccc}
	Criterion & \multicolumn{2}{c}{Threshold 1,000}  & \multicolumn{2}{c}{Threshold 2,000}  & \multicolumn{2}{c}{Difference} \\
	(prevalence) & (day) & (std. dev.) & (day) & (std. dev.) & (days) & (std. dev.) \\
	\hline 
	50 & 148.1 & 6.708 & 172.5 & 7.345 & 24.4 & 9.947 \\
  45 & 151.0 & 8.079 & 175.4 & 7.162 & 24.4 & 10.796 \\
	40 & 153.7 & 8.899 & 177.6 & 7.229 & 23.9 & 11.465 \\
	35 & 156.7 & 9.183 & 181.3 & 7.987 & 24.6 & 12.171 \\
	30 & 161.9 & 9.520 & 182.4 & 6.373 & 20.5 & 11.456 \\
	\hline
	Average & & & & & 23.56 & 11.167 \\
	\hline
	\end{tabular}
	} 
	}
}
\end{table}
\egroup

\subsection{Forecasting}
\label{valid}
This model has been used in Australia in a now-casting mode during the period since 24 March 2020. 
In the simulation timeline, the threshold of 2,000 cases is crossed on day 50, and if this is aligned with 24 March 2020 on the actual timeline, one may see that the incidence along the $90\%$ SD curve starts to reduce from day 59 (aligned with early April 2020), Fig.~3.a of the main manuscript, and the prevalence peak is reached around days 62--65 (aligned with 5--8 April 2020), Fig.~3.b. 

The early projections of the timing of actual incidence and prevalence peaks, as well as three-month ahead forecast of the cumulative incidence in Australia to approach the range of 8,000--10,000 total cases, have shown a good accuracy, validating the model.  Specifically, the agreement between the actual and simulation timelines appears to be the strongest for $90\%$ SD compliance, applied from 24 March 2020 (i.e., primary scenario with 2,000 cases), following a period of weaker compliance between 21 and 24 of March 2020. The predicted cumulative incidence at the end of the suppression period, which maps to the end of June, averages 9,122 cases with 95\% CI [8,898, 9,354], and the range over 20 runs is 8,313 -- 10,090 (see Source Data file).  The actual number of total cases in Australia on 30 June 2020 is reported as 7,834~\cite{wiki-merged}.

Significant levels of compliance have been confirmed by the Citymapper Mobility Index, which collates the usage of the  Citymapper app,  a worldwide public transit app and mapping service which integrates data for all urban modes of transport, for planning public transport, walking, cycling, and micromobility data~\cite{citymapper}. These data allow for approximating the extent of social distancing compliance, showing, by the 26th March 2020, a reduction of 80\% from the normal mobility levels for both Sydney and Melbourne. There was a relatively steep drop in mobility to this level, noting that the number of trips taken by residents of Sydney and Melbourne was around 50\% just five days prior. Since this drop to mid-April, the levels of compliance have remained relatively constant at 80--90\%, peaking at 90\% for both Sydney and Melbourne on the 10th April 2020.  Comparable levels of social distancing were also inferred from the anonymised and aggregated mobile phone location data of several million Australians, provided in early April 2020 by Vodafone, a multinational telecommunications company, to the Australian federal government. These data showed a reduction of 83\% from the normal mobility levels for Sydney, and 82\% for Melbourne~\cite{Vodafone}. A national online survey of 1,420 Australian adults, carried out between 18 and 24 March 2020, found that over the last month 93.4\% of respondents followed at least one of six avoidance-related behaviors~\cite{seale2020covid}. In addition, the ABS survey taken during 1-6 April 2020, showed that during the preceding four weeks, 88.3\% of Australians have been avoiding public spaces (and public events), 98.4\% have been keeping distance from people, and 86.6\% cancelled personal gatherings (e.g., with friends or family)~\cite{ABS4940-1}.

We point out that in Australia, the healthcare sector alone comprises about 6\% the population, with accommodation and food services reaching up to 3.6\%, while transport, postal and warehousing sector occupies 2.6\%,  and electricity, gas, water and Waste services add another 0.6\%~\cite{industry2019}. Thus, assuming that a substantial fraction of employees delivering these essential services cannot work from home, the highest level of social distance compliance would not exceed 90\%.

\vspace*{-2mm}
\section{Comparison of SD compliance levels across several state capitals}	
\label{maps}

\begin{figure}[ht]
\centering 
    \includegraphics[width=0.98\textwidth]{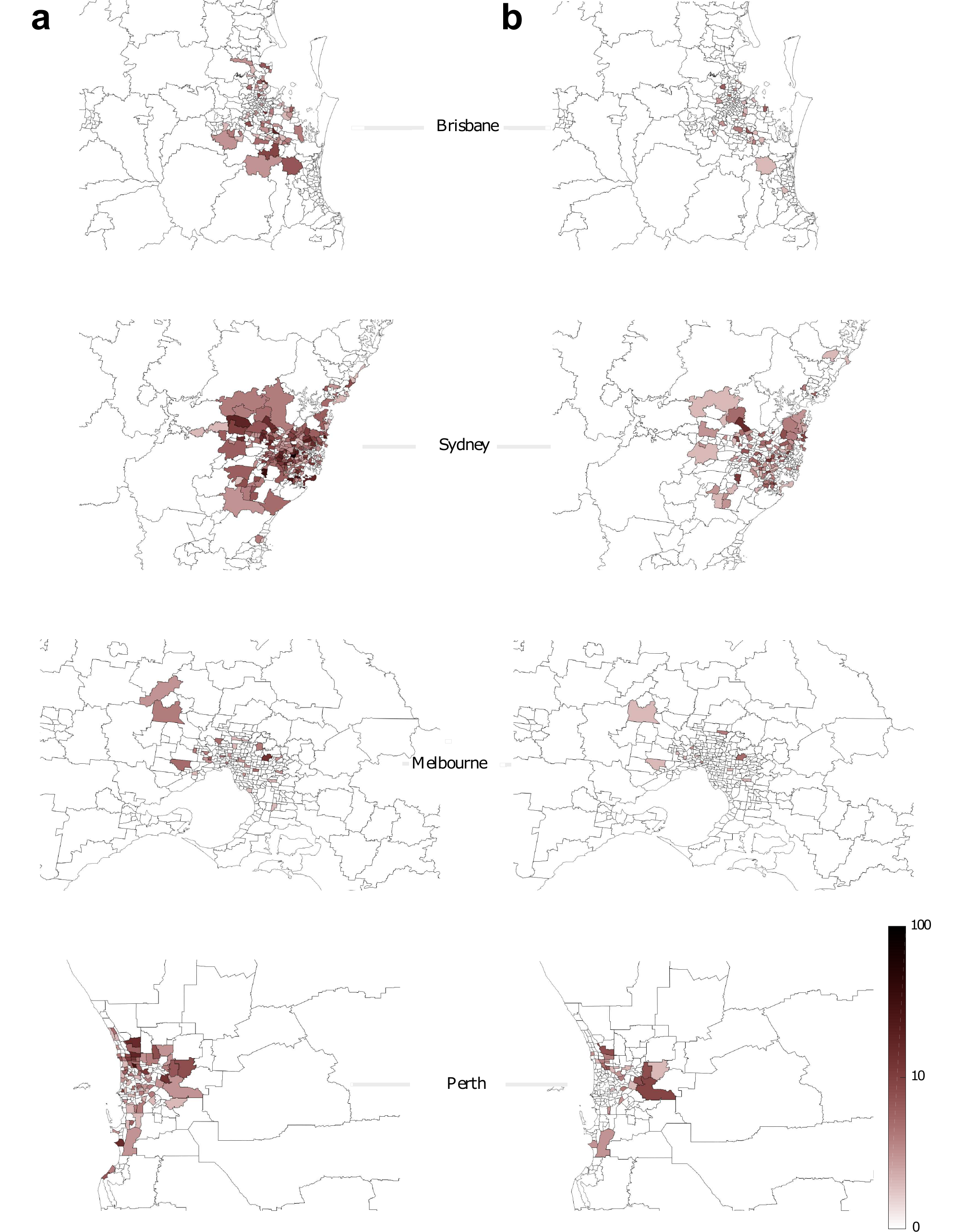}
    \caption{\textbf{Choropleths of four largest Australian Capital Cities.} Prevalence is shown on a log scale at day 60: \textbf{a} 70\% SD compliance, and \textbf{b} 90\% SD compliance.}
    \label{day_60}
\end{figure}

Differences between 70\% and 90\% SD compliance levels are visualised in choropleth maps of four largest Australian Capital Cities: Sydney, Melbourne, Brisbane and Perth (Supplementary Fig.~\ref{day_60}).  These maps contrast prevalence numbers resulting from these two compliance levels at day 60.

\vspace*{-2mm}
\section{Fractions of symptomatic cases across mixing contexts}	
\label{fractions}

Supplementary Table~\ref{tab:fract} summarises {fractions of symptomatic cases} for the considered scenarios, across mixing contexts: households (HH), household clusters (HC), census districts (CD), statistical areas (SA2), working groups (WG), classrooms (CL), grades (GR), schools (SCH).  Notably, a stronger compliance with social distancing, in addition to case isolation and home quarantine, increases the household fraction from 30.48\% (the household fraction under case isolation and home quarantine) to 47.79\% (the household fraction under full lockdown).
This is compensated by the corresponding decreases in the infections acquired at the workplace: from 17.01\% to 6.98\%, as well as in the school environments: from 12.41\% to 6.20\%. 

\bgroup
\begin{table}[h]
	\caption{Average context-dependent fractions of symptomatic cases (over 20 runs, rounded to two decimal places), in \%. NI: no intervention, CI: case isolation, HQ: home quarantine, SC: school closures, SD: social distancing. For NI, CI, HQ and SC: shown at the end of suppression of SC, i.e., after approximately 102 days (including 49 days of suppression). For SD: shown at the end of suppression of SD, i.e., after approximately 143 days (including 91 days of suppression). The contexts include households (HH), household clusters (HC), census districts (CD), statistical areas (SA2), working groups (WG), classrooms (CL), grades (GR), schools (SCH). Source data are provided as Source Data file.}
	\label{tab:fract}
	\vspace{1mm}
	\centering
\resizebox{0.95\textwidth}{!}{%
	{\raggedright
	 \noindent
	{
	\begin{tabular}{lcccccccc}
	Scenario & HH & HC & CD & SA2 & WG & CL & GR & SCH \\
	\hline 
	NI & 17.88	& 17.21	& 25.75	& 13.89	& 20.46	& 1.94	& 1.46	& 1.40 \\
	CI & 28.27	& 15.00	& 17.14	& 9.38	& 17.37	& 5.12	& 3.93	& 3.78 \\
	CI+HQ & 30.48	& 14.31	& 16.65	& 9.14	& 17.01	& 4.95	& 3.80	& 3.66 \\
	CI+HQ+SC & 26.68 & 17.02 & 25.57	& 14.13	& 16.22	& 0.14	& 0.11	& 0.11 \\
	CI+HQ+SD 10\% & 30.61	& 14.88	& 20.87	& 11.35	& 16.54	& 2.32	& 1.76	& 1.67 \\
	CI+HQ+SD 20\% & 32.62	& 14.80	& 20.13	& 10.99	& 15.17	& 2.53	& 1.93	& 1.83 \\
	CI+HQ+SD 30\% & 34.56	& 14.60	& 19.28	& 10.54	& 13.88	& 2.86	& 2.19	& 2.09 \\
	CI+HQ+SD 40\% & 36.63	& 14.34	& 18.36	& 10.04	& 12.55	& 3.20	& 2.49	& 2.38 \\
	CI+HQ+SD 50\% & 38.46	& 14.17	& 17.90	& 9.82	& 11.33	& 3.29	& 2.56	& 2.48 \\
	CI+HQ+SD 60\% & 40.45	& 14.11	& 18.17	& 9.96	& 9.85	& 2.92	& 2.30	& 2.23 \\
	CI+HQ+SD 70\% & 42.45	& 14.07	& 19.01	& 10.34	& 8.24	& 2.32	& 1.82	& 1.75 \\
	CI+HQ+SD 80\% & 44.54	& 13.81	& 18.97	& 10.46	& 7.09	& 1.99	& 1.60	& 1.54 \\
	CI+HQ+SD 90\% & 46.01	& 13.30	& 18.33	& 9.97	& 6.85	& 2.17	& 1.75	& 1.63 \\
  CI+HQ+SD 100\% & 47.79	& 12.78	& 16.99	& 9.25 & 6.98	& 2.45	& 1.90	& 1.85 \\
	\hline
	\end{tabular}
	} 
	}
}
\end{table}
\egroup

\clearpage

\bibliographystyle{naturemag}

\end{document}